\documentclass[10pt]{article}
\usepackage[dvips]{graphics}
\usepackage{amsmath}
\usepackage{latexsym}
\usepackage{amsfonts}

\begin{document}

\def\eq{\begin{equation}}
\def\en{\end{equation}}

\begin{center}
\vskip 3cm
{\large \bf Non-Ohmic hopping transport in a-YSi: from isotropic to directed percolation}

\vskip 0.8cm

 F. Ladieu$^{*}$

 {\it Laboratoire Pierre S\"ue, DRECAM, CEA-Saclay, 91191 Gif / Yvette
CEDEX, FRANCE}

 D. L'H\^ote$^{\dagger}$ and R. Tourbot

 {\it Service de Physique de l'Etat Condens\'e, DRECAM, CEA-Saclay, 91191 Gif / Yvette
 CEDEX, FRANCE}

\vskip 0.6cm

\end{center}

  (Received                                  

\vskip 1.2cm

 \par

\vskip 1cm

Electrical transport has been investigated in
 amorphous Y$_{0.19}$Si$_{0.81}$, from $30 \, \hbox{mK}$ to room temperature.
 Below $2\, \hbox{K}$, the conductance $G$ exhibits Efros-Shklovskii behavior 
${G} \sim \exp[-(T_0/T)^{1/2}]$ at zero electric field, 
where conduction is expected to occur along a very sinuous path 
(isotropic percolation). 
The non-linear {\it I-V} characteristics are systematically studied 
up to very high fields, for which the conductance no longer depends on $T$
and for which the current paths 
are expected to be almost straight (directed percolation). 
 We show that the contributions of electronic and sample heating to those
non-linearities  
are negligible. Then, we show that the conductance dependence as a function
of low electric fields ($F/T < 5000 \, \hbox{V}\ \hbox{m}^{-1} \ \hbox{K}^{-1}$)
 is given by $G(F,T)=G(0,T)\times \exp(-eFL/k_B T)$. The order of magnitude (5-10 nm) 
 and the $T$ dependence ($\sim T^{-1/2}$) of $L$ agrees with theoretical predictions.
 From the $T_0$ value and the length characterizing the intermediate field regime, we
 extract an estimate of the dielectric constant of our system. 
 The very high electric field data do not agree with the prediction $I(F) \sim \exp[-(F_0/F)^{\gamma \prime}]$ with $\gamma \prime \,= 1/2$ : we find a $F$ dependence of $\gamma \prime$ which could be partly due to tunneling across the mobility edge. In the intermediate electric field domain,
 we claim that our data evidence both the enhancement of the hopping probability with the field {\it and the influence of the straightening of the paths}. The latter effect is due to the gradual transition from isotropic to directed percolation and depends essentially on the statistical properties of the ``returns'' i.e. of the segments of the paths where the current flows {\it against} the electrical force. The critical 
exponent of this returns contribution which, up to now was unknown both theoretically and experimentally, is found to be $\beta = 1.15 \pm 0.10$. An estimation of the returns length is also given.
\vskip 0.4cm

\pagebreak
\begin{center}
{\bf I. INTRODUCTION}
\end{center}

The electrical transport in disordered insulators such as doped semiconductors
 or amorphous systems, has been widely investigated for several decades as a convenient way to probe the properties of Anderson 
 insulators \cite{Anderson58}. At finite temperature $T$,
 the conduction results a priori from all the electronic 
hops between any localized states $i$ and 
$j$ separated by a distance $r_{ij}$ and a difference in energy
$E_{ij}$. Starting from the probability of such a hop 
$p_{ij} \propto \exp{-({r_{ij}}/{\xi} + {E_{ij}}/{k_{B}T})}$ 
(with $\xi$ the localization length and $k_{B}$ the Boltzmann constant), Mott 
\cite{Mott69} introduced the key concept of variable-range hopping (VRH) stating that 
transport is dominated by {\it the most probable hops} whose characteristic  length $r_{m} \propto T^{-1/(d+1)}$ and energy 
$E_{m} \propto  T^{d/(d+1)}$ depend on $T$. As a result, the conductance $G$ is predicted to follow
$$G  = G_0 \exp{\Bigl\lbrack -{ \Bigl( {\frac{T_{0}}{T}} \Bigr) }^{\! \gamma} \,  \Bigr\rbrack} \mskip 300mu  (1)$$

\hskip -5.8mm where $G_{0}$ and $T_{0}$ are material dependent constants, and
 $\gamma = {1}/({d+1})$ depends on the dimensionality $d$ of the sample.  Later, the percolation theory 
 has been used to derive more rigorously Eq. (1) \cite{Ambegaokar71} \cite{Pollak72}. In addition
Efros and Shklovskii \cite{Shklovskii74} found that electron interactions result in a ``soft'' Coulomb gap at the Fermi level, leading to $\gamma = {{1}/{2}}$. It is currently admitted now \cite{Aharony92} that Mott and Efros-Shklovskii laws are limiting cases of VRH. Both of them were observed experimentally
as their relevance depends
on the relative magnitudes of the hopping energy and Coulomb gap.

\begin{center} 
{\bf A. Electric field effects in hopping transport }
\end{center}

The case of very high
fields $F$, ${eF\xi}/{(k_{B}T)} \gg 1$ ($e$ = electron charge) was studied through different 
theoretical approaches \cite{Shklovskii72}-\cite{Vandermeer82} which all predict that the current $I$ should no longer depend on the temperature 
(activationless hopping) and behaves according to

$$I = I_0 \exp{\Bigl\lbrack -{\Bigl({\frac{F_{0}}{F}}\Bigr)}^{\! \gamma\prime } \, \Bigr\rbrack }  \mskip 300mu (2)$$

\hskip -5.8mm where $I_0$ and $F_{0}$ are constants and
 $\gamma\prime = \gamma$ (see Eq. (1)).
In spite of their similarity, Eqs. (1) and (2) correspond to 
completely different current path topologies. 
Eq. (2) is obtained by considering the shortest hops along which the electrostatic energy 
gain overcomes the energy fluctuations due to disorder \cite{Hill71}, \cite{Shklovskii72}. Thus in the very high field case,
 the hops occur in the direction of the field
 ({\it directed} percolation). On the contrary, for $F \to 0$, 
 the current flows through
 a network of random impedances along {\it very sinuous} paths 
whose returns and meanders are such that any hop more resistive 
than the VRH prediction is forbidden. A quantitative description of these paths can be obtained by using the {\it isotropic} percolation theory \cite{Shklovskii75} \cite{Feng91}. 

 The intermediate field 
case ${eF\xi}/({k_{B}T}) < 1$, where neither $T$ nor $F$ can be neglected
is the most difficult one to handle theoretically, and the various models lead to different predictions that we summarize in the general equation giving the current 
$$I(F,T) = I_1 \exp{\Bigl\lbrace
 - \Bigl( {\frac{T_0}{T}} \Bigr)^{\gamma} \ 
 \Bigl\lbrack 
1-A{\frac{(eF\xi)^{\alpha}} 
{(k_{B}T)^{\alpha \prime}}} +B \Bigl({\frac{eF\xi} 
{k_{B}T}} \Bigr)^{\beta} \  
\Bigr\rbrack \Bigr\rbrace } \mskip 80mu (3)$$

\hskip -5.8mm where $A$, $B$, $\alpha$, $\alpha \prime$ and $\beta$ are parameters, $A$ and $B$
being positive.
 The first term $A(eF\xi)^{\alpha}/(k_{B}T)^{\alpha'}$ expresses the enhancement of the hopping probability when 
$F$ grows. Depending on the authors, $\alpha$ and $\alpha \prime$ range from 
$\alpha=\alpha'={1}/({\nu+1})$ \cite{Shklovskii76} (where $\nu \simeq 0.88$ for $d=3$ is the critical index of the correlation radius) to $\alpha=\alpha'=1$ \cite{Pollak76}, \cite{Hill71}, and even $\alpha = 2$ and $\alpha'={9}/{4}$ for $d=3$ \cite{Apsley75}, \cite{Vandermeer82}. The discrepancies between these results come from the fact that the net current $I_{ij}$ between two sites may increase exponentially with $F$ or be insensitive to $F$, depending on the relative positions of the site 
energies. The second term $B({eF\xi}/( 
{k_{B}T}))^{\beta}$ was predicted {\it only} by B\"ottger,
 Bryksin {\it et al.} \cite{Bottger80}-\cite{Bottger86} who addressed the additional
 problem of the ``returns'' , i.e. of the segments
 of the paths where the current flows {\it against} 
the electrical force. Their model states that the length of these returns decreases 
{\it gradually} when $F$ increases. We show in this paper  
that our data evidence the existence of both terms of Eq. (3) with
 $\alpha = \alpha' = 1$ and $\beta = 1.15 \pm 0.10$.

\begin{center}
{\bf B. Theories of hopping transport at intermediate fields}
\end{center}

We recall here some theoretical points which are 
 important for the interpretation of our data.
The very sinuous current paths at $F \to 0$ 
 can be pictured by using 
the ``nodes-links-blobs'' concept \cite{Shklovskii75} \cite{Feng91}, derived from the isotropic percolation theory :
 at small scales the current path 
consists of a fractal-like network of connected conductances up to the 
``blob'' scale ${\cal L}_{p}$, and the blobs are the links of a homogeneous network.
 In the linear regime, the 
conductances inside a blob are much larger than the key
 conductances $G_{key} \sim \exp{- \left({T_{0}}/{T}\right)^{\gamma}}$ 
lying at the end of each blob. $G_{key}$ sets the overall sample 
conductivity since at scales larger than ${\cal L}_{p}$ the system is 
homogeneous. For the intermediate field case, 
where neither $T$ nor $F$ can be neglected, three main 
effects have to be considered : 

(i) for a site $i$ belonging to the percolation path, the density of 
sites $j$ yielding a non negligible conductance $G_{ij}$ is modified by $F$;

(ii) a field-induced rearrangement of the charge 
along the current paths can take place : the local chemical potential
 $\mu_{i}$ can fluctuate widely around the mean value given by $eFr_{i}$;

(iii) the lengths of the ``returns'', i.e. the segments where the
 current has to go against the electrical force, decrease when $F$ 
increases. 

Most authors disregarded (iii) and focused upon (i) and (ii), yielding
 predictions summarized by Eq. (3) with $B \equiv 0$. 
 However, they gave
 diverging predictions depending on the relative influence given to (i) and (ii).
 For example, 
Pollak and Riess \cite{Pollak76} find 
$\alpha=\alpha'=1$ and $A=3/32$ in a model where (i) plays the central role, and the correlations 
between nearest neighbors conductances are taken into account. They 
find that (ii) 
is quite negligible ($\delta \mu_{i} \le k_{B}T$).
 This contradicts Shklovskii model
 \cite{Shklovskii76} 
($A \simeq 1$ and $\alpha=\alpha'={1}/({1+\nu}) \simeq 0.53$ for $d=3$) which 
emphasizes (ii) ($\delta \mu_{i} \gg k_{B}T$) by 
concentrating almost {\it all} the electrical potential drops existing at the blob scale
 ${\cal L}_{p}$ upon the key conductance. The
 disagreement between these predictions  
was never completely explained \cite{Talamantes87} even if numerical 
simulations by Levin and Shklovskii \cite{Levin84} suggested that the Pollak and Riess result holds in usual experimental 
conditions $({T_{0}}/{T})^{{1}/{(d+1)}} \le 70$ while Shklovskii 
result would hold in the opposite case, i.e. at exceedingly low 
temperatures. 
 Two other groups 
found different results : Apsley {\it et al.} \cite{Apsley75}
 and Van der Meer \cite{Vandermeer82} 
obtained $\alpha = 2$, $\alpha \prime = 9/4$, but their methods were
 criticized since the former did not take into account the directed percolation requirement, while the latter uses dimensional ``invariants'' which were shown later to be $F$ dependent \cite{Bottger82}.

Taking (iii) into account\cite{Bottger80}\cite{Bottger82}\cite{Bottger84} prevents the current from increasing too fast : the corresponding second term $B({eF\xi}/ 
{k_{B}T})^{\beta}$ in Eq. (3) reduces the exponential increase of $I$ due to 
$-A{(eF\xi)^{\alpha}}/ 
{(k_{B}T)^{\alpha \prime}}$.
 The physical reason is that the returns act as ``bottlenecks'' for the current.
 The problem being exceedingly complicated, the precise values of $\alpha, \beta, A, B$ were
 found to depend on the various methods used by B\"ottger, Bryksin {\it et al.} 
 The last and more accurate results were obtained numerically \cite{Bottger84}\cite{Bottger85} 
\cite{Bottger86} : $\alpha = \alpha \prime = 1$, $A={1}/{6}$ , $B \simeq 0.02$, $1/\beta \simeq 1.1$ 
 (this latter parameter is the critical exponent of
 the returns length $\Lambda$, see below Eq. (12)). The theoretical value of $\beta$ {\it is given within an error of $50\%$ due to numerical uncertainties.} Yet, the {\it precise} value of $\beta$
 is of major importance since for
 $\beta < \alpha$, the differential conductivity $\sigma = {\partial I}/{\partial V}$ decreases at low fields to reach  
an absolute 
minimum at a finite $F$. On the contrary, in the case $\beta > \alpha$, 
$\sigma$ increases at low fields and has no absolute minimum at 
finite field.

 The above mentioned theories disregard 
 the possible $F$-dependence of the carrier density due to trapping in dead ends of paths, as well as  the possible  $F$-dependence of the 
 localization length $\xi$. Fortunately, both effects should be negligible in our case. Indeed, 
{\it statistically} as many holes as electrons 
become trapped when $F$ increases, as in our case the density of 
states is symmetric around the Fermi level 
(parabolic Coulomb gap)\cite{Talamantes92}. The $\xi(F)$ dependence should be 
negligible \cite{Kirkpatrick86} as long as $E_{F} \gg eF\xi$ where
 $E_{F}$ is the kinetic energy at the Fermi level: this is the case in our sample\cite{note1}.

\begin{center}
 {\bf C. The experimental situation}
\end{center}

On the experimental side, {\it I-V} non linearities were investigated both 
in amorphous materials \cite{Morgan71}-\cite{Popescu98}
and in doped crystalline semiconductors \cite{Redfield75}-\cite{Stephanyi97}.
In most of these works, the authors focussed either on very high fields or on intermediate
fields.
For intermediate fields, the data 
\cite{Marshall73} \cite{Elliott74}  \cite{Aleshin87} \cite{Redfield75}-\cite{Timchenko89} \cite{Grannan92}
 were analyzed using the electric field dependence predicted by Hill 
\cite{Hill71}, Pollak and Riess \cite{Pollak76}, and Shklovskii \cite{Shklovskii76}
\vskip -0.1 cm
$$\ln{\bigl(G(F,T) \bigr)} = \ln{\bigl( G(0,T) \bigr)} + {\frac{eFL}{k_{B}T}} = \ln{( G_0)} - \Bigl({\frac{T_{0}}{T}} \Bigr)^{\gamma} + {\frac{eFL } {k_{B}T}}  \mskip 60mu  (4)$$ 
\vskip 0.1 cm

\hskip -5.8mm where $L$ is a length related to the hopping distance $r_m = (\xi/2)(T_{0}/T)^{\gamma}$. Hill\cite{Hill71} and Pollak and Riess\cite{Pollak76} predict
$L=C.(\xi/2)(T_{0}/T)^{\gamma}$  with $C=0.75$ or $0.18$, while Shklovskii
predicts $L \propto T^{-(1+ \nu) \gamma}$ \cite{Shklovskii76} with $\nu=0.88$.
Eq. (4) results from Eq. (3) by neglecting the returns term ($B=0$) and by assuming that the $F$ and $T$ 
dependences of $G$ and $I$ are close to each other. 
The experimental values of $L$ and their $T$ dependence often disagree with the theoretical
predictions \cite{Aleshin87} \cite{Zabrodskii80}-\cite{Ionov87}\cite{Kenny89}.
The claim by Wang {\it et al.} \cite{Wang90} that the non-linearities in Ge:Ga were due to carriers heating \cite{Cooper87} rather than to the preferential hopping along the electric field which leads to Eqs. (3) or (4) triggered hot-electron analyses of the data \cite{Stephanyi97}-\cite{Perrin97}. They show that 
the relative contribution of carriers heating and electric field effects has to be considered
at very low temperature.
In very high electric fields, the expected activationless conduction (see Eq. (2)) has been observed by several authors \cite{Nair77}-\cite{Dvurechenskii88} \cite{Timchenko89} \cite{Tremblay89} \cite{VanderHeijden92}. 
The equality of the VRH 
and activationless exponents ($\gamma$ and $\gamma \prime$ in Eqs. (1) and (2)) has been checked for $\gamma=1/2$ in several systems  \cite{Dvurechenskii88} \cite{Tremblay89}, but $\gamma \prime=1/4$ and $\gamma=1/2$ has been found in amorphous GeCu\cite{Aleshin87}.

In this work we systematically investigate the whole set of non linearities
 ranging from the linear VRH regime at $F \to 0$, to the activationless conduction at
very high fields : in our low temperature domain ($0.03$ K $ \le T \le 1.3$ K), $F/T$ 
varies from $1.5  \times  10^2$ $\hbox{V} \, \hbox{m}^{-1} \, \hbox{K}^{-1}$ to $1.4  \times  10^6$ $\hbox{V} \, \hbox{m}^{-1} \, \hbox{K}^{-1}$.
 Our samples are made of amorphous  Y$_{0.19}$Si$_{0.81}$.
 In comparison to most of the experimental works performed
in this low temperature range ( $<$ 1 K) \cite{Rosenbaum80} \cite{Tremblay89}-\cite{Stephanyi97}, we go to higher values of 
$F$ and $F/T$, thus allowing an investigation of both the 
intermediate and the very high field regimes. 
At very low temperatures, this has been done by Rosenbaum 
{\it et al.} on Si:P \cite{Rosenbaum80}, however the authors
 found an abnormally large value of $L$. At higher temperatures, 
these two regimes were investigated
 for {\it a}-GeCu at 1.6 K $<$ $T$ $<$ 110 K \cite{Aleshin87} and 
ZnSe at 1.6 K $<$ $T$ $<$ 4.2 K \cite{Timchenko89}. The first paper
 raised the question of unexpected exponents values (in the $L$ vs. $T$ and
 $\ln(G)$ vs. $F$ dependences) while the second one found a 
puzzling decrease of $L$ when $F$ increases. Clearly, new data
are needed. In this work, we test the validity of Eqs. (1)-(4) and show
that they are in good agreement with our data 
for intermediate fields. In particular we find
 $L \simeq C.r_m$ \cite{Hill71}\cite{Pollak76}. We also check that 
the contribution of heating effects to the non-linearities is negligible.
 Our main result
is that Eq. (3) with $\alpha$ = $\alpha \prime$ is more suitable than 
Eq. (4) to fit the data : this allows us to estimate the size of the
returns and the value of the related critical exponent $\beta$\cite{Bottger80}-\cite{Bottger86}.
 Finally, we show that in the activationless region,
the $I$ vs $F$ dependence is more intricate than the law given by Eq. (2).

 The paper is organized as follows. In section II we describe the 
experimental setup and method.  
The conductance as a function of the temperature in the 
region $F \to 0$ is presented in section III. In section IV, we consider 
the very high field case. Having extracted the physics of the two extreme $F$ regimes, 
we finally turn in section V
to the intermediate field case. Finally, we summarize our main results in section VI.
  In appendix A we show that heating mechanisms cannot explain the $I-V$
non-linearities, while appendix B is devoted to the extraction of the localization length
from the data.

\begin{center} 
{\bf II. EXPERIMENT}
\end{center}

As shown in Fig. 1, each amorphous Y$_{x}$Si$_{1-x}$ sample studied ($x \simeq$ 0.19) is deposited on 
a sapphire substrate on which two interdigited gold electrodes were evaporated and
 etched. The spacing between the electrodes is $l = 128 \, \mu$m. 
The Y$_{x}$Si$_{1-x}$ layers were obtained by Argon plasma sputtering on the sapphire substrate 
(cooled at 77 K to prevent Y or Si aggregates formation),
 using a Y$_{x}$Si$_{1-x}$ source. The thickness of the layers is 
$9 \, \mu$m, much larger than any hopping length, hence electrical transport is three dimensional. 
Previous studies of electrical transport in 
 such Y$_{x}$Si$_{1-x}$ samples 
\cite{Boucher88}-\cite{Ladieu96}, have shown that a small variation of $x$ 
 allows to cross the metal-insulator transition : above 
$x \simeq 0.22$ the samples are metallic; while
 for lower values of $x$, the samples exhibit an insulating behavior,
 i.e. a divergence of the resistance for $T \to 0$, in agreement with
the VRH predictions \cite{Pichard90}-\cite{Specht95}, except for the 
very weakly insulating samples at the lowest 
temperatures \cite{Ladieu96}. In the present work $x \simeq 0.19$ and the samples are 
strongly insulating, exhibiting VRH below $T\simeq 2 \,$ K. Since the two samples we used behaved similarly, we report only the 
data of one of them. 

As shown in Fig. 1, a great attention was paid to the thermalization of the 
sample because of a possible heating interpretation of
the {\it I-V} curves non-linearities. As explained in Appendix A, we measured the thermal conductances involved in our sample and concluded that heating effects were irrelevant in our sample: neither heating of the whole sample nor carrier heating can explain the {\it I-V} non linearities reported here. 

Electrical contacts were made by ultrasonic soldering of two gold wires on the 
evaporated electrodes. The data 
were obtained by setting a given voltage $V = Fl$ on the sample, and by
waiting for a time $\tau$ before measuring the current $I$. We paid the greatest attention to choosing $\tau$ large enough
 to let the current settle, e.g. we took $\tau (T<800 \, \hbox{mK}) \ge 100 \, \hbox{s}$. The temperature was registered for each $I(V)$ point, and its stability proved to be better than $1 \% $. 
 The coaxial cables used were previously tested alone (in an open circuit, at all temperatures) to check that their leakage current
was negligible. Let us note that it is unlikely that the 
gold-YSi electrical contact resistance plays a role since the contact area is very
 large and in previous similar works \cite{Specht95}, 
it was shown to play no role.
 The fact that the temperature dependence $G(0,T)$ we observe follows a VRH law
 (see next section) is an additional indication that the contact resistance is negligible.
 Finally, the symmetry of the {\it I-V} curves with respect to current and
voltage reversing was checked.

\begin{center} 
{\bf III. BEHAVIOR AT $F \to 0$}
\end{center}

The {\it I-V} characteristics are shown in 
Fig. 2 where, for clarity, we report only the curves for $18$ different temperatures out of the $36$ ones we measured. As shown in the 
inset, the resistance ($R$) - temperature law is simply activated from room temperature
 to $\simeq 10 \, \hbox{K}$ : we find $R=R_{1} \hbox{exp}(-T_1/T)$ 
with $T_{1} = 40 \pm 4\, \hbox{K}$ and $R_{1} \simeq 88 \pm 2 \, \Omega$. 
For $T<8 \, \hbox{K}$ the divergence 
of the resistance when $T \to 0$ becomes {\it weaker} (see inset of Fig. 2), which is 
a standard indication that VRH takes place at low 
temperature \cite{Shklovskii74}.
Let us note that simple activation for $T \ge 10 \, \hbox{K}$ is usually interpreted as 
due to thermal excitation from the Fermi level $E_{F}$ 
to the mobility edge $E_{Mob}$ 
where electronic states are delocalized; thus
 $k_{B}T_{1} = E_{Mob}-E_{F}$. 

To study the temperature dependence of the conductance at $F \to 0$, we
plotted our data in the ($F/T$, $\ln{(V/I)}$) plane (see Fig. 3).
 The $G(F \to 0)$ values were obtained from an extrapolation of the curves 
towards low fields. To allow for a precise extrapolation, 
 this procedure was restricted to the $0.4 \, \hbox{K} < T < 1.3 \, \hbox{K}$ cases.
 The extrapolated values were obtained from a linear fit of 
 the first points in each curve (lower $F/T$ values). We checked that those values did not 
 depend significantly on the number of points selected in the fit.
  The expected VRH temperature-conductance law is given by Eq. (1) 
 where $\gamma = {1}/{2}$ or $\gamma = {1}/{4}$.
 Discriminating between these 
exponents is hardly achieved by looking for a straight line
in the $\ln(G(F \to 0))$ versus $T^{-\gamma}$ plot, as the
lines always {\it seem} straight. The comparison between the 
different exponents $\gamma$ is improved by plotting a normalized
value of $T^{-\gamma}$, i.e. $(T^{-\gamma}-T^{-\gamma}_{Min})/(T^{-\gamma}_{Max}-T^{-\gamma}_{Min})$
where $T_{Min}$ and $T_{Max}$ are the two extreme
 values of $T$ selected for the plot.
In Fig. 4, we see that $\gamma=1/2$ is favored in comparison with 
$\gamma=1/4$. A fit of $\ln(R(F \to 0))$ vs. $T$ using Eq. (1) 
in which $G_0$ and $T_0$ are free parameters yields a residue (measured minus fitted value) shown in the inset of Fig. 4 : clearly, there is
a correlation between the residue and $T$ only for $\gamma = 1/4$.
 The normalized $\chi^2$ per point is 1.1 for $\gamma = 1/2$,
and 9.0 for $\gamma = 1/4$. If $\gamma$ is a parameter of the fit,
 we find $\gamma = 0.47 \pm .02$. We thus conclude that $\gamma = 1/2$, and
using this value in the fit, we get 
$T_{0} = 257 \pm 1.5 \, \hbox{K}$ and $R_{0} = 1/G_0 = 92 \pm 5 \, \Omega$. 
 Following Efros and Shklovskii, 
$$k_B T_{0} = \frac{2.9 e^{2}}{4 \pi \epsilon_{0} \epsilon_{r} \xi} \mskip 350mu (5)$$

\hskip -5.8mm where  $\epsilon_{0} \epsilon_{r}$ is the dielectric constant 
of the system. We shall see in Appendix B that the non-linearities analysis together
with Eq. (5)
allow to extract an information on $\xi$ and $\epsilon_r$ separately.  

\begin{center} 
{\bf IV. RESULTS AT VERY HIGH FIELDS}
\end{center}

In the very high fields case, ${e F \xi}/({k_{B}T}) \gg 1$, 
the theory \cite{Shklovskii72} states that the current
results from
hops between sites $i$ and $j$, such that $E_{ij} = eFr_{ij}$. For these activationless hops, 
$I_{ij} \propto \exp{(-2{r_{ij}}/{\xi})}$, which, using a maximization procedure
 leads to Eq. (2) with\cite{Shklovskii72} 

$$ F_0 = N_2 \frac{k_{B}T_{0}}{e\xi} \ \mskip 350mu (6)$$

\hskip -5.8mm where $N_{2}$ is a numerical constant which can be calculated within the 
directed percolation theory. For Mott's VRH ($\gamma = {1}/{4}$) Pollak and Riess 
obtain $N_{2} = 4.8$. For the Efros-Shlovskii VRH 
($\gamma = {1}/{2}$), $N_{2}$ has not been calculated to our knowledge.

Our lowest temperature data should belong to the very high field
domain, as the two {\it I-V} curves corresponding to $T=29 \, \hbox{mK}$ and $T=49 \, \hbox{mK}$ 
are identical. Indeed, we shall see that the low and intermediate
field data analysis lead to $\xi \approx 7 \, \hbox{nm}$ (see Appendix B),
hence ${e F \xi}/({k_{B}T}) \ge 50 $ for $T=29 \, \hbox{mK}$. 
 The $29 \, \hbox{mK}$ data can 
be rather well fitted by using Eq. (2) with $\gamma \prime = 1/2$,
 $I_0$ and $F_0$ being free parameters. It yields
 $F_{0} = (3.8 \pm .05) \times 10^{7} \, \hbox{V/m}$, hence $4 \le N_{2} \le 11$ by using 
the upper and lower values of $\xi$. The normalized $\chi^2$ per point is however 
 290. Thus, in spite of the reasonable extracted $F_0$ and $N_2$ 
values, the relevance of the fit must be questioned.
 Indeed, if we fit the whole $I(F, 29 \, \hbox{mK})$ data using Eq. (2),
 $\gamma \prime$ being a free parameter,
we get $\gamma \prime = 0.65 \pm 0.01$ and a $\chi^2$ per point of 94. 

To investigate the problem, we give in Fig. 5 the $F$ dependence of the
 effective exponent $\gamma_{eff}$ defined by 
$$\gamma_{eff} = -{\frac 
{\partial \ln{
\left( \frac{\partial \ln{I}}{\partial \ln{F}} \right)}}
{\partial \ln{F}}
} . \mskip 280mu (7)$$

\hskip -5.8mm If $I(F)$ is given by Eq. (2), then $\gamma_{eff} = \gamma \prime$ .
For $T > 200 \, mK$, $\gamma_{eff}$
is always negative, while for $T < 200 \, \hbox{mK}$, $\gamma_{eff}$
it is negative for the lower $F$ values. This can be related
to the fact that in the intermediate field region, the $I(F)$
dependence should obey
Eq. (3) for which the expected $\gamma_{eff}$ is $F$-dependent.
For the lowest temperatures, $\gamma_{eff}$ becomes positive when 
$F$ increases, but there is no saturation at
$\gamma_{eff} = 0.5$. Instead, a maximum is reached at a value between 0.5 and 1, even if a convergence of $\gamma_{eff}$
towards $1/2$ might exist at fields larger than 
$4 \times 10^4 \, \hbox{V/m}$.

We suggest that these values of $\gamma_{eff}$ between 0.5 and 1 are due to the fact that 
the very high field conduction results from an interplay between activationless hopping
(for which $\gamma_{eff}=1/2$) and {\it tunneling across the mobility edge}. Indeed, the latter process leads to 
the simple activated law \cite{Nebel92}, \cite{Sze81}
 
$$I \sim 
  \exp {\Bigl\lbrack -(\frac{4}{3eF})\sqrt{\frac{2m(k_{B}T_{1})^{3}} {\hbar^{2}} }\Bigr\rbrack} =  \exp \Bigl\lbrack{- \frac{F_{1}}{F}}\Bigr\rbrack\mskip 210mu (8)$$

\hskip -5.8mm where $k_{B}T_{1}$ is the energy difference between the Fermi level 
and the mobility edge. Since $T_{1} \simeq 40 \, \hbox{K}$ (see Sec. III), 
$F_{1} = 1.4 \times 10^{6}\, \hbox{V/m}$. The fit of the
$I(F, 29 \, \hbox{mK})$ data for the interval
$F > 2.8 \times 10^4 \, \hbox{V/m}$
where $\gamma_{eff} > 0.5$,
 using  $I=I_1 \exp{(-{F_{1}}/{F})}$, yields a normalized $\chi^2$ per point of
40 and  $F_{1} 
= (5.5 \pm 0.05) \, 10^{5} \, \hbox{V/m}$, which is only a factor 2.5 below the calculated value. We cannot interpret the $\gamma_{eff}(F)$ curves 
 more precisely since, to our knowledge, there exists no theory 
taking into account both the disorder in the localized band states and the tunneling 
through the Mobility Edge.

It is interesting to note that if activationless 
hopping was the only transport mechanism in the very high field regime, the critical value $({F}/{T})_{c}$ at which the transition from Eq. (3) to Eq. (2) 
occurs would be temperature independent : this can be readily seen by equating the rhs of
Eqs. (2) and (3) at ${F}/{T}=({F}/{T})_{c}$ and disregarding pre-exponential factors.
If we define, for any given $T$ the experimental critical field value $F_c$ by 
$I(F_{c},T) = k\,I(F_{c},29 \ \hbox{mK})$, with $k=1.1$, we find that in our sample, $({F}/{T})_{c}$ is clearly T-dependent:
it increases by a factor 3.5 when the temperature decreases from 413 mK to 74 mK.

\begin{center} 
{\bf V. RESULTS AT INTERMEDIATE FIELDS}
\end{center}

 We can see on Fig. 3 that our experimental points line up
in almost parallel curves in the ($\ln{(R)}$, $F/T$)
plane. Clearly, they are not really straight lines, hence
 Eq. (4) does not hold with precision. However, in the spirit of the 
majority of previous works, we first use Eq.(4) and extract the length $L(T)$ defined  
by $\ln{[G(F)/G(F \to 0)]} = eFL(T)/(k_{B}T)$ for $F$ just above the linear regime. We show in a second step that Eq. (3) is much more relevant to account for our data. 

\vskip 0.5cm
\begin{center} 
{\bf  A. Extraction of $L(T)$ using Eq. (4)}
\end{center}

 The slopes of the
 $\ln{(R)}$ vs. $F/T$ curves for $F \rightarrow   0$ were
 extracted for the different temperatures where it 
was possible, i.e. $0.4 \, \hbox{K} < T < 1.3 \mskip 5mu \hbox{K}$ 
(see Fig. 3). A linear fit yielded a normalized
$\chi^2$ per point close to 1 only for $F/T < 5000 \mskip 5mu \hbox{V} \mskip 3mu \hbox{m}^{-1} \mskip 3mu \hbox{K}^{-1}$. 
We restricted the $L$ extraction to this interval where the curves are linear.  
 The result
 is shown in Fig. 6. 
A fit of these points using the 
law $L=L_{0}  \mskip 5mu (T_0/T)^\zeta$ where $L_0$ and 
$\zeta$ are free parameters gives $\zeta \simeq 0.65$ with a
 normalized $\chi^2$ per point of 0.7. 
Constraining the fit with $\zeta = 0.5$ 
leads to $L_{0} = 
(0.52 \mskip 5mu \pm \mskip 5mu 0.01) \mskip 5mu \hbox{nm}$ and a normalized $\chi^2$ per point of 1.1 :
the resulting curve is shown in Fig. 6. 
 As a consequence, we can state that the Shklovskii prediction
 \cite{Shklovskii76} $\zeta = 2 \gamma = 1$ does not hold for our system, while
 the predictions of Hill \cite{Hill71} and Pollak and Riess\cite{Pollak76}, $L \propto r_m$
 leading to $\zeta = \gamma = 1/2$,
 are in agreement with our results.
 It is interesting to note that the values of $L(T)$ remain roughly the same
 if they are extracted from linear fits of the {\it whole} $\ln(R)$ vs. $F/T$ curves,
 although these fits yield a normalized $\chi^2$ per point varying between 3 and 250.
 The localization length $\xi$ extracted
 from $L_0=C.(\xi /2)$ with the Pollak and Riess value $C=0.17$ 
\cite{Pollak76}
 yields $\xi \simeq 6 \, \hbox{nm}$.
 We shall come to this point more in detail later.    

\vskip 0.25cm
\begin{center} 
{\bf B. Extraction of the parameters of Eq. (3)}
\end{center}

We turn to the main result of our study, namely the relevance of the 
 B\"ottger {\it et al.} prediction \cite{Bottger85} concerning the ``returns''
 and summarized in Eq. (3).

\vskip 0.25cm

B.1. { $F/T$ is the {\it unique} relevant parameter} 

We define, so as to test Eq. (3), 
$$\Psi (F,T)  = \sqrt{\frac{T}{T_{0}}} \ln{ 
\left(\frac{I(F,T)}{I_{E.S.}(F,T)} \right) } 
{\hbox{ , with } }\ I_{E.S.}(F,T)= \frac{V}{R_{0}} \exp{ [-(\frac{T_{0}}{T})^{1/2} ] }
 \mskip 15mu (9)$$

\hskip -0.58 cm where $V=Fl$, $R_{0} = 92 \, \Omega$ and $T_{0} = 257 \, K$ (see Sec. III).
It is easy to show that if Eq. (3) holds in what concerns the $G(T,F)$ dependence
(instead of $I(T,F)$), $\Psi$ is just equal to $A(eF\xi)^{\alpha}/(k_B T)^{\alpha \prime}-B(eF\xi/k_B T)^{\beta}$. 
We have chosen to perform our analysis on $G(T,F)=I(T,F)/(Fl)$ rather than on $I(T,F)$ 
 because
 all the above depicted theories do not allow to recover a linear regime
 when $F \to 0$ : in this case, Eq. (3) gives 
$I(F \to 0) = I_1 \exp{[-(T_0/T)^\gamma]}$ which is not consistent with 
Eq. (1). 
Note that this difficulty has been ignored
 in almost all experimental works dealing with the intermediate fields
 case, as they focus upon $G(F)$ rather than
 on $I(F)$. 
 Numerically, our choice is justified
by the fact that the exponential $I(F)$ dependence in Eq. (3)
 is much faster than the contribution of the denominator of $G(F) \sim I(F)/F$ (see also 
\cite{note2}).

To perform our analysis of the intermediate field case, we suppressed the 
data points corresponding to the very high field case (see Sec. IV).
This was done by selecting 7164 $I(F,T)$ points among the whole set of
8189 points, on the criterion $I(F,T)/I_{VHF}(F) - 1 \ge 10\%$, where $I_{VHF}(F)$
 is the very high field current defined by
 $ I_{VHF}(F)=I(F,T=29 \, mK)$. 
The $10\%$ criterion is rather arbitrary, but we checked that 
 our results do not depend on its precise value,
 provided it remains above $5\%$. Then, $\Psi$ is calculated for each of these
7164 points. 
As shown on Fig. 7, $\Psi(F,T)$ {\it is a universal function of} ${F}/{T}$ over 
more than $3$ orders of magnitude, while the  $7164$ experimental points
 correspond to $36$ different temperatures ranging from 49 mK to 1.29 K. 

The fact that $\Psi(F,T)$ depends only on $F/T$ implies $\alpha=\alpha'$, which definitely shows that 
both Apsley \cite{Apsley75} and Van der Meer \cite{Vandermeer82} models 
are not relevant for 
our sample. Moreover, the log-log slope of $\Psi({F}/{T})$ {\it at the lowest} 
$F/T$ values is very close to $1$ and clearly larger
than ${1}/{(\nu + 1)} \simeq 0.53$ which means that Shklovskii prediction 
\cite{Shklovskii76} does not hold. This is not surprising since 
in our experiment $({T_{0}}/{T})^{1/2} \le 70$ which is a domain where
the numerical work of Levin and Shklovskii \cite{Levin84} leads to 
 $\alpha = 1$. However our data are poorly fitted by the Hill \cite{Hill71} and the Pollak and Riess\cite{Pollak76}
 prediction $\Psi \sim F/T$ since the log-log slope of $\Psi({F}/{T})$ 
decreases as $F/T$ increases. More generally, a fit of $\Psi(F/T)$ with a unique power law $(F/T)^{\alpha''}$ gives a poor agreement whatever $\alpha''$, especially because of the points in the region $F/T \ge 3 \times 10^{4} \, \hbox{V K}^{-1}\hbox{m}^{-1}$. The only remaining prediction is thus 
the one of B\"ottger {\it et al.} where $\alpha=1$ and $\beta$ is unknown theoretically. In the next section this prediction will be {\it assumed} and the best value of $\beta$ will be sought. Quite interestingly, the fit of $\Psi$ we obtain in the next section is, 
by far, much better than all the ``reasonably simple'' fits we tried: for example  
 attempting a polynomial fit with $\Sigma_{i=1}^{m} a_{i}(F/T)^{i}$ gives a poor agreement 
even if $m$ is as large as $5$. 

\vskip 0.25cm

{ B.2. Extraction of the critical exponent $\beta$ assuming $\alpha = 1$.}

 We thus have to fit the $\Psi(F/T)$ points using the function 
$$\Gamma {\frac{F}{T}} - \Delta \bigl({\frac{F}{T}} \bigr)^\beta  \mskip 300mu (10)$$ 

\hskip -5.8mm with $\Gamma, \Delta, \beta$ as free parameters. Since the $\Psi(F/T)$ values extends 
over several orders of magnitude, we perform the fit on $\ln{( \Psi)}$, i.e.
 we request that a {\it relative} error 
${\cal E}$ per point is minimal, with $\cal E$ defined by  
$$ {\cal E}^2 =  \frac{1}{7164 s^2}\sum_{j=1,7164} 
(1-x) \Bigl[ \ln{\Bigl( \frac{\Psi \bigl( (F/T)_j \bigr)}{\Gamma (F/T)_j - \Delta (F/T)_j^\beta} \Bigr)} \Bigr]^2$$

 $$ + x \Bigl[\ln{\Bigl( \frac{ \Gamma (F/T)_{j} - \Psi \bigl( (F/T)_j \bigr) }{\Delta (F/T)_j^\beta}  \Bigr)} \Bigr]^2 \\\\\\\, ,
 \mskip 190mu (11)$$

\hskip -5.8mm where $x$ is a weight to be chosen in the $(0,1)$ interval and 
$s^{2} =$ $ \left< (\ln{(\Psi )} - <\ln{(\Psi )}>)^{2} \right> \simeq (0.08)^2$ 
is the variance of $\ln({\Psi})$ drawn from the data. 
Note that the first term in the rhs of Eq. (11) is the relative error per point 
on $\Psi$ while 
the second one is the relative error per point on the ``B\"ottger second term'' of Eq. (3), 
$\Delta (F/T)^{\beta}$. 
 The minimization of those two terms has {\it opposite effects} on the optimal value of $\beta$.
 Indeed keeping only the first one (i.e. setting $x=0$) leads to $\beta =  1.38 \pm 0.03$ and yields a good agreement between the $\Psi ((F/T)_{j})$ points and the fitted curve, but leads to a discrepancy
between $\Gamma (F/T)_{j} - \Psi \bigl( (F/T)_j \bigr)$ and 
$\Delta (F/T)_j^\beta$ which becomes clearly too large when 
$F/T < 5 \times 10^{4} \, \hbox{V} \ \hbox{K}^{-1} \ \hbox{m}^{-1}$.
 Conversely, setting $x=1$ leads to 
$\beta = 1.03 \pm 0.03$, with
 $\Gamma (F/T)_{j} - \Psi \bigl( (F/T)_j \bigr) \simeq \Delta (F/T)_j^\beta$, but
 a discrepancy appears between the $\Psi \bigl( (F/T)_j \bigr)$ points and the fitted curve, especially at high $F/T$ values. 
 Keeping $x$ in the $(0.05 , 0.95)$ interval leads to optimal values $\beta(x)$ in the $(1.05 , 1.15)$ range, i.e. to an uncertainty on $\beta$ not very much larger than the one obtained for a given $x$. We finally keep the larger error bar on $\beta$ and find :
 $\beta = 1.15 \pm 0.1$, 
$\Gamma = (1.25 \pm 0.5) \times 10^{-5}\, \hbox{K m/V}$ and $\Delta = 1.4 
\times 10^{-6} \hbox{(m K/V)}^{\beta}$ (${\Delta^{{1}/{\beta}}}/{\Gamma} = 0.70 \pm 0.07$).
 The solid line 
on Figs. 7 and 8 is the fit resulting from
 $\left( \beta=1.15; \Gamma = 1.25\times 10^{-5} \hbox{K} \, \hbox{m} \, \hbox{V}^{-1}; \Delta = 1.4 
\times 10^{-6} (\hbox{K} \, \hbox{m} \, \hbox{V}^{-1})^{\beta} \right)$ : the agreement with the data is good (${\cal E}(x=0.5) =0.72$) and the only systematic deviation occurs at the very 
few highest $F/T$ values where 
the transition to the very high fields regime occurs.

 The comparison of Eqs. (3) and (10) gives $\Gamma = A{e\xi}/{k_{B}}$. 
 From $\Gamma = 1.25 \times 10^{-5}\ \hbox{m K/V}$ we deduce $A\xi = 1.1\, \hbox{nm}$. Unfortunately, the precise value of 
$A$ is not known in our case of Efros Shklovskii's hopping
 ($\gamma= 1/2$). As shown in Appendix B, using $T_{0}=257 \, \hbox{K}$, $A\xi = 1.1\, \hbox{nm}$ as well 
as considerations about the dielectric constant $\epsilon_{r}$ we conclude 
that $\xi \approx 7 \, \hbox{nm}$. 
We cannot be more precise due to the various unknown numerical factors involved in the 
predictions we used, but we note that these values of $\xi$ compare 
favorably to previous results obtained on $Y_{x}Si_{1-x}$ samples much closer to the metal-insulator transition \cite{Specht95}, \cite{Ladieu96} where larger values of $\xi$ were found (a few tens of nm).

We compare now the analysis performed just above with the one 
 carried out using Eq. (4) which yielded $L=L_0 \, (T_0/T)^{1/2}$ with
 $L_0 = 0.52 \, \hbox{nm}$, and $\xi \simeq 6 \, \hbox{nm}$,
 using $L_0 = C \, (\xi /2)$ and $C=0.18$ (see Sec. B.). 
Neglecting the weak difference between the $G(T,F)$ and the $I(T,F)$
dependences, the comparison of Eqs. (3) and (4) leads to 
$L_0 = A \xi$ if the additional term $B(eF\xi/(k_B T))^{\beta}$ in Eq. (3)
 is {\it assumed not to change} $L_0$.
 Clearly, we have a discrepancy by a factor 2 between $L_0$ and $A \xi =1.1 \, \hbox{nm}$ :
 this was checked to come from the $B(eF\xi/(k_B T))^{\beta}$ term in Eq. (3).
 Surprisingly both analysis
 yield roughly the same $\xi \approx 6 - 7 \, \hbox{nm}$ for $A=1/6$ and $C=0.18$.

\vskip 0.25cm

{ B.3. Extraction of the return length $\Lambda$.}

 We now turn to the analysis of the
``returns'' size.
As explained above, $\beta$ and $\Delta$ give an information about 
the typical length $\Lambda$ of the returns which decreases gradually 
as $F$ increases. Indeed, according to B\"ottger
 {\it et al.} \cite{Bottger84}, if $\rho = \ln{[I(F,T)/I_{1} ]}$ 
 ($\rho$ is the argument of the exponential in Eq. (3)), 
 $\Lambda$ is given by 
$$\Lambda(\rho) \simeq  2 \delta r_{m} 
\Bigl(\frac{\rho_{c}}{\rho - \rho_{c}} \Bigr)^{\frac{1}{\beta}} 
\Bigl(\frac{\rho_{m} - \rho }{\rho_{c}}\Bigr)^{\mu} \mskip 150mu (12)$$

\hskip -5.8 mm where $\mu = 1.0 \pm 0.3$ and $\delta \simeq 0.25$ are numerical constants
 \cite{Bottger84}, $\rho_{m}$ is the directed percolation threshold, and $\rho_c =  ({T_{0}}/{T})^{\gamma}$.
 In the linear regime it is found that $\rho(F\to 0) \simeq \rho_{c} + 1$
 \cite{Shklovskii75}, yielding a finite value of 
$\Lambda (F \to 0)$.
 Increasing $F$ up to the intermediate field region
 leads to a $\rho$ increase, thus to a $\Lambda$ decrease. Increasing $F$ further, $\rho$ reaches 
$\rho_{m}$ and $\Lambda$ vanishes indicating that the current paths are directed. 
The maximum length $\Lambda_{max}$ of the returns is thus obtained just at the 
onset of the non linear regime, where we can put in Eq. (12)
 $\rho =({T_{0}}/{T})^{\gamma} + 1$ and 
${(\rho_{m} - \rho)}/{\rho_{c}} \simeq {(\rho_{c} - \rho_{m})}/{\rho_{c}} = N_{4}$ with $N_{4} \simeq 0.07$ as predicted in \cite{Bottger84}. 
Hence we get 

$$\Lambda_{max} = \xi \delta N_{4}^{\mu} 
\left( \frac{T_{0}}{T} \right)^{\gamma (1+\frac{1}{\beta}) } = {\Delta^{\frac{1}{\beta}} }
\frac{k_{B}}{e} \left( \frac{T_{0}}{T} \right)^{\gamma (1+\frac{1}{\beta}) } \mskip 120mu (13)$$

\hskip -5.8mm where the second equality is obtained by using the B\"ottger {\it et al.} prediction 
$\Delta = ({N_{4}^{\mu} \delta e \xi}/{k_{B}})^{\beta}$ \cite{Bottger84}. 
Using Eq. (15), we can compare $\Lambda_{max}$ to the blob 
length ${\cal L}_{p}= \xi (T_{0}/T)^{\gamma (1+\nu ) }$ \cite{Shklovskii75} :  
$$\frac{\Lambda_{max}}{{\cal L}_{p}} = 
\frac{\Delta^{\frac{1}{\beta}} k_{B}}{e \xi} \left( \frac{T_{0}}{T} \right)^{\gamma (\frac{1}{\beta}-\nu ) } \simeq \frac{\Delta^{\frac{1}{\beta}} k_{B}}{e \xi} 
\mskip 150mu (14)$$

\hskip -5.8mm where the second equality comes from the fact that in our case, due to the
 experimental value of $\beta$, $\Lambda_{max}$ {\it and}
 ${\cal L}_{p}$ {\it have critical exponents very close to each other}
 since ${1}/{\beta}-\nu$ is in the (-0.08, 0.07) interval.
 We thus find that ${\Lambda_{max}(T)}/{{\cal L}_{p}(T)}$ {\it does not depend on the temperature}. By inserting ${k_{B}}/{(e \xi )} = {A}/{\Gamma}$ in Eq. (15), we find 
${\Lambda_{max}(T)}/{{\cal L}_{p}(T)}= A{\Delta^{\frac{1}{\beta}}}/{\Gamma} = (0.70 \pm 0.07)A\ $, using our fit result 
${\Delta^{\frac{1}{\beta}}}/{\Gamma} = 0.70 \pm 0.07$. 
 Hence, ${\Lambda_{max}(T)}/{{\cal L}_{p}(T)}$ {\it is not much smaller than} $1$ which is the highest possible theoretical value.
 We thus conclude that in our sample the importance of the returns is strong.

This importance of the returns must also play a role on the onset 
of the non linear behavior 
which occurs at a field $F_{lim}(T)$. Indeed, according to 
B\"ottger {\it et al.}  
$F_{lim}/T = k_B /e\Lambda_{max}$ \cite{Bottger84}. According to Pollak
{\it et al.}, $F_{lim}/T = k_{B}/er_{m}$
 with $r_{m} = ( \xi /2 )(T_0 /T)^{\gamma}$ the hopping length. 
 In our case, this latter prediction amounts to $F_{lim}/T \ge 1000 \, \hbox{V K} \, \hbox{m}^{-1}$ for
 $T\ge 500 \, \hbox{mK}$ (for lower  temperatures we cannot go to $F \to 0$ because of too low currents).
 We clearly see on 
Figs. 7 and 8 that the non linear fit obtained above extends down to 
$F/T \simeq 150 \, \hbox{V} \ \hbox{K}^{-1} \ \hbox{m}^{-1}$. This
 is one order of magnitude smaller than the Pollak {\it et al.} prediction and in quite good agreement 
with the B\"ottger {et al.} result.
 We thus conclude
 that the $F_{lim}$ behavior we find is consistent with 
the above derived results on the importance of returns.

\begin{center} 
{\bf VI. CONCLUSION}
\end{center}

Our study of the electric field effects in 
variable-range hopping transport
for amorphous Y$_{0.19}$Si$_{0.81}$ below 2 K exhibits several important features.
 First, we find that the length $L$ characterizing the 
 intermediate field regime has the order of magnitude 
 and $T$ dependence ($\simeq T^{-\gamma}$, with $\gamma=1/2$)
 which is expected in the VRH models of Hill \cite{Hill71} or Pollak and Riess \cite{Pollak76}
 stating that $L \simeq r_m $. Even 
 analyzing our data in the framework of the B\"ottger, Bryksin {\it et al.}
 predictions (Eq. (3)) this result remains true, as the addition of the 
 $B(eF\xi /(k_B T))^\beta$ term in Eq. (3) does not change the order of magnitude
 of this characteristic length.
 Second, our most important result is that Eq. (3) is much more
relevant than Eq. (4) to analyze our data, and {\it this shows the 
importance of the ``returns'' in the percolation paths of VRH}.
Furthermore, we were able to extract an information on the length
of these returns from our experimental results.
Indeed, our data indicate that the critical exponent
of the returns contribution $1/\beta$ is very close to the
one of the blob length : the returns represent an appreciable
$T$-independent fraction of the whole percolation paths
lengths at intermediate fields. 
 Third, our very high field data do not follow the expected activationless
law given by Eq. (2). This could be due to the onset of 
tunneling across the mobility edge whose interplay with 
 activationless hopping has not been theoretically studied yet.

\vskip 0.5cm
{\bf ACKNOWLEDGEMENTS}\par
We want to thank J.P. Bouchaud, J.L. Pichard and  M. Sanquer for fruitful 
discussions. Many thanks also to Y. Imry for explaining us
 Ref. \cite{Imry82}.

\pagebreak

\begin{center} 
{\bf APPENDIX A: IRRELEVANCE OF THE SAMPLE OR CARRIERS HEATING MECHANISM}
\end{center}

In this Appendix, we first describe the experimental setup allowing thermal conductances measurements and then show that heating is irrelevant to account for our {\it I-V} non linearities.

 The sapphire substrate was hold between four sapphire 
balls which represent a negligible thermal conductance. The thermalization was
 realised by ultrasonically bonding 43 gold wires ($50 \,\mu \hbox{m}$ in diameter, a few millimeters long) between
 the cryostat (``Heat sink'' on Fig. 1) and a $7 \, \hbox{mm}^2$ gold pad schematized on the left of the substrate on Fig. 1. The resulting calculated thermal resistance 
between the gold pad and the cryostat is smaller than $10^{4} \, \hbox{K/W}$ at $T=50 \, \hbox{mK}$
 and $10^{3} \, \hbox{K/W}$ at $T=500 \, \hbox{mK}$, leading to a temperature 
 difference lower than $0.5 \, \hbox{mK}$ between those two points in our experiments. 
 To measure the thermal resistance 
between the cryostat and the sapphire substrate or the sample,
we used the $1 \, \hbox{M}\Omega$ heating
resistor chip shown on the right up corner of Fig. 1. This $0.01 \, \hbox{mm}^3$ 
device is held {\it only} by two Al wires ($25 \, \mu \hbox{m}$ in diameter, 5 mm long) which are
 superconducting below $\simeq$ 1 K. At low temperature,
 their thermal conductance is thus negligible and the electrical power
 dissipated in the resistor flows to the substrate through the unique gold wire ($25 \, \mu \hbox{m}$
 in diameter, 3 mm long) ultrasonically bonded between
 the resistor and the gold pad evaporated on the sapphire substrate (on the right in Fig. 1). By measuring the temperature of the sample when a given power is injected in the resistor,
 this system gives an over-estimation of the thermal resistance between the 
 sapphire substrate and the cryostat. It has the advantage of being reversible in comparison with
a resistor deposited on the sapphire substrate, as the gold wire can be easily removed.

We show now that heating phenomena are irrelevant to account for 
the {\it I-V} non linearities. We first focus on the possible
 heating of the whole sample together with its sapphire substrate, and then turn to ``electronic heating''.
 
\begin{center}
{\bf A. Heating of the whole sample and substrate}
\end{center}

The measurement of the thermal resistance ${\cal R}_{th}$ between the cryostat on one hand and the sample with the sapphire substrate on the other hand was 
carried out as follows. A constant voltage $V=Fl$ was applied to the
Y$_{x}$Si$_{1-x}$ sample, leading to a current $I(F,T)$ depending
 on the cryostat temperature $T$. A given electrical power
 $1 \, \hbox{nW} \le {\cal P} \le 100 \, \hbox{nW}$
 was then dissipated in the 
small $1 \, \hbox{M}\Omega$ resistor.
 This power flowed to the heat sink through the substrate whose temperature
 was then increased by $\delta T$, leading to an increase $\delta I$ of the current.   We checked that $\delta I$ 
was proportional to ${\cal P}$.
 Assuming for a while that the {\it I-V}
non linearities do not result from sample heating, 
we extracted, from the data of Fig. 2,
 $\delta T$ from $\delta I$ and deduced 
${\cal R}_{th} = {\delta T}/{\cal P}$. 
 Fig. 9 gives the resulting
${\cal R}_{th}$ as a function of $T$. From it, we get
 ${\cal R}_{th}(74 \, \hbox{mK}) \simeq 10 \, \hbox{mK/nW}$, while we
 can see on Fig. 2 that the
 $I(F, 74 \, \hbox{mK})$ and $I(F, 124 \, \hbox{mK})$ curves begin to merge for $IV 
\simeq 0.04 \, \hbox{nW}$ which corresponds to $\delta T \simeq 
0.4 \, \hbox{mK}$ much lower than the $50 \, \hbox{mK}$ separating these two $I(F)$
 characteristics. This confirms the above assumption of irrelevance of 
sample heating; and it can be easily shown to be true at any temperature.
 Let us note that ${\cal R}_{th}$ exhibits a ${\cal R}_{th} \sim T^{-3}$ behavior which characterizes the Kapitza thermal resistance at the boundary between two materials. In our case, they are the sapphire substrate
and the gold pad thermally connected to the cryostat.

\begin{center}
{\bf B. Electronic heating}
\end{center}

In this section, we show that the non-linearities of our {\it I-V} curves 
cannot be ascribed to a heating effect as found by Wang {\it et al.}
 for NTD sensors \cite{Wang90}. 
Assuming the validity of such a model for our sample, the power
 $P=IV=IFl$ injected in the YSi ``electron bath'' would
increase its temperature to a value $T_e$ larger than the 
cryostat temperature $T$, leading to a YSi electrical resistance

$$R(T,F) = R_0 \, \exp{((T_0/T_e)^{\gamma})}\, ,  \mskip 260mu (\hbox{A1})$$

\hskip -5.8mm with $T_e$ given by

$$P + P_p =IV=g(T_e^{\eta} - T^{\eta}) \mskip 300mu (\hbox{A2})$$

\hskip -5.8mm where $g$ and $\eta$ are parameters characterizing the thermal conductance
between the electron bath and the cryostat, and $P_p$ is the parasitic power injected
 in the sample due to the limited rf shielding etc. As we have excluded
a possible heating effect of the sapphire substrate (see previous section), 
the thermal resistance to consider is either the Kapitza one at
the boundary between the YSi sample and the sapphire,
 or the thermal resistance due to electron-phonon coupling in the YSi
 itself. In the first case $\eta = 4$\cite{Schwartz89}, 
while in the second one, $\eta = 5$ to $6$ \cite{Wang90}\cite{Stephanyi97}-\cite{Perrin97}\cite{Cooper87}. We investigated the
experimental values of $dP/dT_e$, which would be equal to
$g{\eta}T_e^{\eta-1}$, hence would not depend on $T$ and $P_p$.
Within the heating model, $T_e$ can be extracted for each ({\it I,V}) point using Eq. (A1).
Fig. 10 shows $dP/dT_e$ as a function of $T_e$. The fact that most
 of the curves depend on $T$ is a strong argument against the 
heating model. However, we can see a trend towards a $T$ independence for low
$T$ values. As indicated by the straight lines corresponding to power
 laws, these low $T$ curves are not compatible with the heating model
 because they correspond to $\eta$
 values which depend on $T_e$ and may be much larger than $4$ to $6$.  
 Finally, we note that in materials close to {\it a-}YSi,
 the electron-phonon coupling parameter $g$ was found 
experimentally to be of the order of $10^3 \mskip 4mu \hbox{W K}^{-6} \hbox{cm}^{-3}$ (in NbSi, with $\eta = 6$) \cite{Dumoulin96} or $200 \mskip 4mu \hbox{W K}^{-5} \hbox{cm}^{-3}$ (in AuGe,
 with  $\eta = 5$) \cite{Cooper87}. As a consequence, for our $ 7.2 \times 10^{-5} \, \hbox{cm}^3$ sample, we
expect $dP/dT_e \simeq 0.4 \mskip 4mu \hbox{T}_e^5 \mskip 6mu \hbox{W/K}$ or $dP/dT_e
 \simeq 0.07   \mskip 4mu T_e^4 \mskip 6mu \hbox{W/K}$ which is several
 orders of magnitude larger than what we find (see Fig. 10). If we assume
 that the possible heating effect is due to the YSi-sapphire Kapitza resistance, 
we can use the experimental $g \simeq (1-10) \times 10^{-3}
 \mskip 4mu \hbox{W} \ \hbox{K}^{-3} \hbox{cm}^{-2}$ values for sapphire-metal
 interfaces\cite{Yvon96}, which lead to  $dP/dT_e$ values
 of the order of $10^{-3} \mskip 5mu T_e^3 \mskip 6mu \hbox{W/K}$, again
 several orders of magnitude larger than our experimental values. 
Yet, those very large discrepancies guarantee that the heating effects are
negligible. In comparison to data from authors who see heating effects,
 this can be 
explained by : i) our very resistive sample, ii) the low electron-phonon and Kapitza
 resistances due to the geometry of the sample.

\pagebreak
\begin{center} 
{\bf APPENDIX B: ESTIMATION OF $\bf \xi$ AND $\epsilon_r$ FROM $\bf T_{0}$ AND $\bf \Gamma$.}
\end{center}

  The detailed calculations of $A$, $B$, $\alpha$, $\alpha \prime$ and $\beta$ 
were performed by B\"ottger {\it et al.} assuming Mott VRH ($\gamma = {1}/({d+1})$) 
while we have $\gamma = 1/2$. Thus our fit result $A \xi= 1.1 \ \hbox{nm}$ cannot be 
used straightforwardly to get $\xi$. In this appendix, we first try to estimate
 in which interval  $A$ must lie in the case of $\gamma= 1/2$ and then we use the value of $T_{0}$ as an additional experimental constraint on $\xi$.

It appears that the exponents $\alpha$, $\alpha'$ and $\beta$
 should remain unchanged when going from $\gamma = 1/(d+1)$ to $\gamma = 1/2$. This can be readily seen for $\alpha = \alpha' = 1$ in the framework of Pollak's calculations \cite{Pollak76}. 
Moreover, the fact that $\alpha = 1$ for $\gamma = 1/2$ is confirmed by many analyses of experimental results
 using Eq. (4) 
\cite{Elliott74} \cite{Aleshin87} \cite{Redfield75} \cite{Ionov87} \cite{Kenny89} \cite{ChenGang89} \cite{Timchenko89} \cite{Grannan92}; while, in what concerns $\alpha \prime$ the experimental
situation is unclear (see Sec. I).
 However, the prefactors $A$ and $B$ are likely to be changed by the presence of a Coulomb gap.
 In this case, we can state at least $A \le {1}/{2}$, since $A = {1}/{2}$ comes from the replacement of 
${E_{m}}/{(k_{B}T)} \simeq \frac{1}{2}(T_0/T)^{\gamma}$ by ${E_{m}}/{(k_{B}T)} - {eFr_{m}}/{(k_{B}T)}$ in Eq. (1). Such a substitution obviously  
overestimates $I$ in the presence of an electric field since it neglects both the influence of returns and the insensitivity of 
a large number of pair currents with respect to $F$. From $A \le {1}/{2}$ we deduce $\xi \ge 2.2 \, \hbox{nm}$.
 Using the B\"ottger, Bryksin {\it et al.} value \cite{Bottger84} \cite{Bottger85} \cite{Bottger86}
 $A$ = 1/6, we find $\xi = 6.6$ nm which, using Eq. (5) gives $\epsilon_r = 29$, while $\xi \ge 2.2 \, \hbox{nm}$
leads to $\epsilon_r \le 86$.

 According to Imry {\it et al.} \cite{Imry82},
 one expects $\epsilon_{r} = N_{1} ({\xi}/{\lambda_{T.F.}})^{2}$ where 
$\lambda_{T.F.}$ is the Thomas Fermi screening length and $N_{1} = 1$ 
according to Abrahams and Lee \cite{Abrahams86}. Thus, using Eq. (5),
 $\lambda_{T.F.} \simeq 0.08  A^{-3/2} \, \hbox{nm}$ which leads to $\lambda_{T.F.} = 1.2 \, \hbox{nm}$ for $A=1/6$ and $\lambda_{T.F.} \ge 0.23 \, \hbox{nm}$ for $A \le \ {1}/{2}$.
In crystalline 
metals with a concentration $n$ of one electron per atom we have typically
 $\lambda_{T.F.} \simeq 0.06 \, \hbox{nm}$.
 Here we expect 
$n$ to be of the order of $0.1$ and since the standard screening theory yields 
$\lambda_{T.F.} \propto n^{-1/6}$, the value of $0.23 \, \hbox{nm}$ for our sample is plausible. However, we cannot exclude that the upper value of $1.2 \, \hbox{nm}$ is plausible too. Indeed, it was theoretically found \cite{Nakhmedov87} 
that close to the transition, electronic diffusion is considerably 
lowered, which should reflect in a lowering of screening, i.e. in an (unknown) increase of 
$\lambda_{T.F.}$. Moreover, if as recently suggested \cite{Ortuno98}, 
many body effects come into play in Efros-Shklovskii VRH, one expect both a reduction 
of the $2.9$ factor in Eq. (5) and a change of the $A$ value.
 Finally, considering 
all these unknown effects, $A \xi$ is found to have the correct order of magnitude and we 
estimate that $\xi \approx 7 \, \hbox{nm}$.
 
 The determination
 of the dielectric constant $\epsilon_r$ of the system is a very interesting 
working direction for the future due to the fundamental interest of $\epsilon_r$
in localization and MIT studies. This is also a strong argument 
in favor of a precise theoretical determination of the
numerical parameters in Eq. (3). Even if those parameters are not 
completely known, the {\it relative} evolutions of $\epsilon_r$
as a function of the dopant concentration, magnetic field, etc. 
should be reachable with the $\epsilon_r$ extraction method we used.

\pagebreak

$^*$ Electronic address : ladieu@drecam.cea.fr , Phone : (33) (0)169088558,  Fax: (33) (0)169086923
 
$^{\dagger}$ Electronic address : lhote@amoco.saclay.cea.fr , Phone : (33) (0)169083015

\pagebreak 

\begin{center}
{\bf Figure Captions}
\end{center}

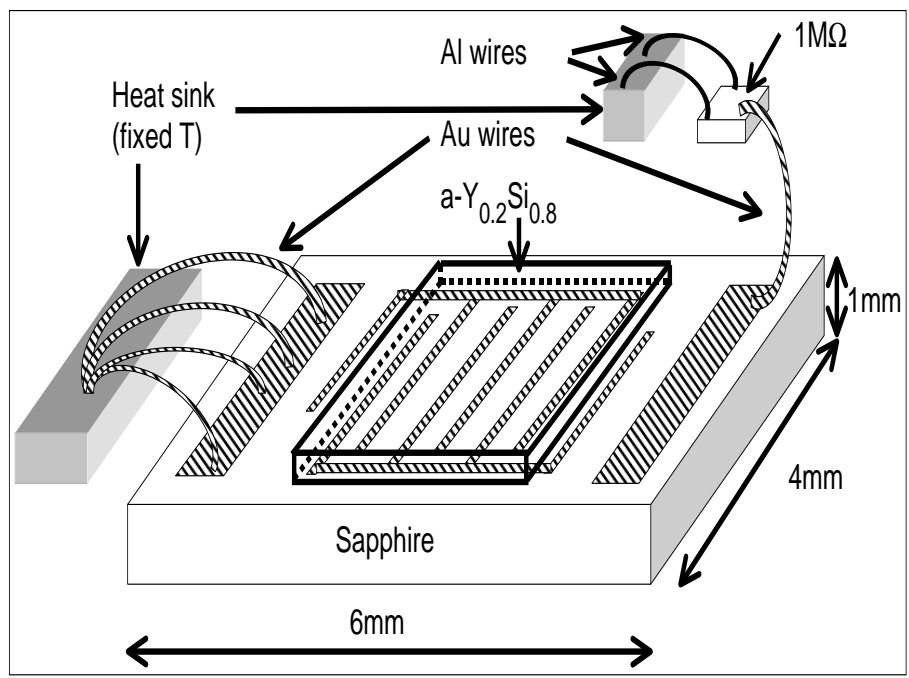
FIG. 1. Experimental setup: the $9 \, \mu \hbox{m}$ thick 
Y$_{0.19}$Si$_{0.81}$ layer is deposited onto a sapphire substrate with two interdigited gold electrodes. Thermalization of the sample was ensured by 43 gold wires (4 represented) bonded on the left gold pad. Upper right corner :
 small $1 \, \hbox{M}\Omega$ chip heater allowing thermal resistance measurements at low temperature (see Appendix A).

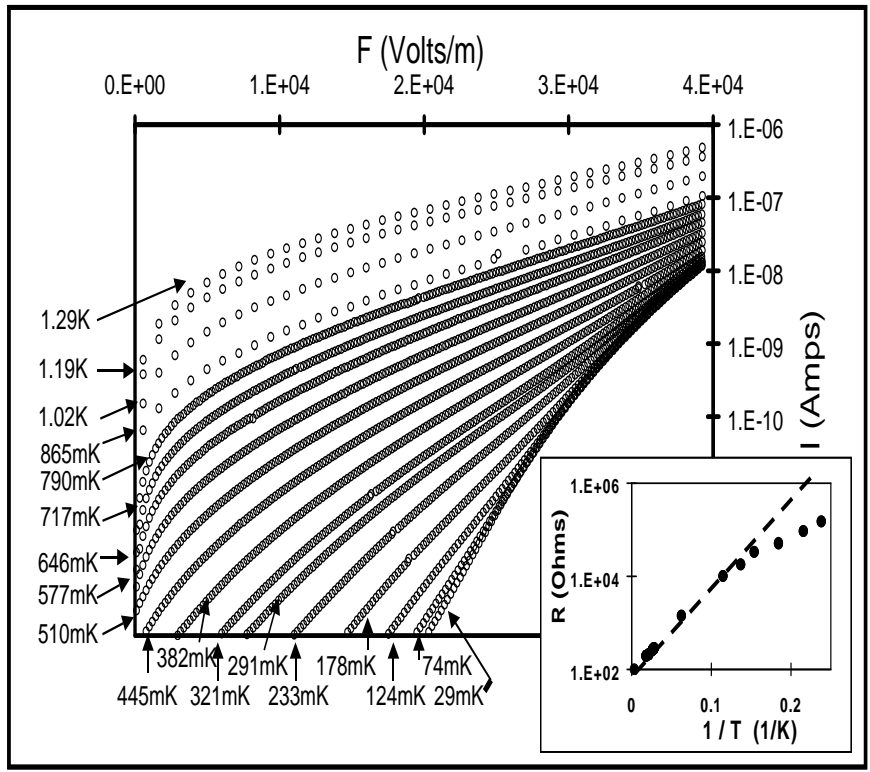
FIG. 2. Current as a function of electric field 
at various temperatures below 1.3 K : for clarity only $18$ out of the $36$ different curves were represented. Inset:  $R(F \to 0)$ vs. $1/T$ (dots), for temperatures 
above 4 $K$. The dashed line corresponds to an activated law
 with a characteristic temperature of 40 K.

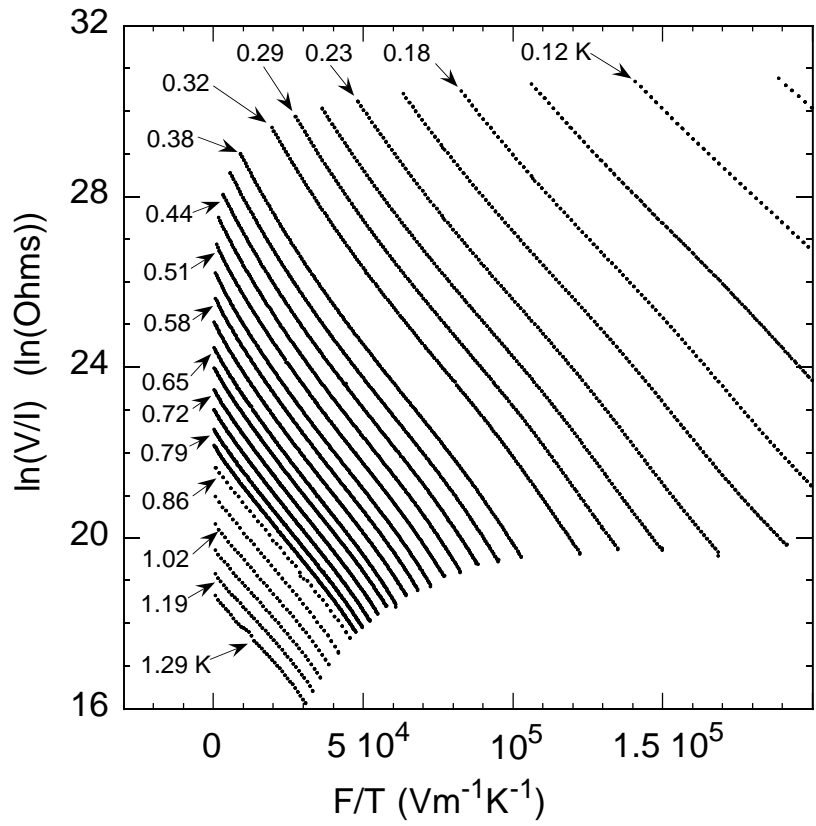
FIG. 3. Logarithm of the resistance as a function the electric
 field to temperature ratio, for 14 temperatures among the 28 measured ones.
 
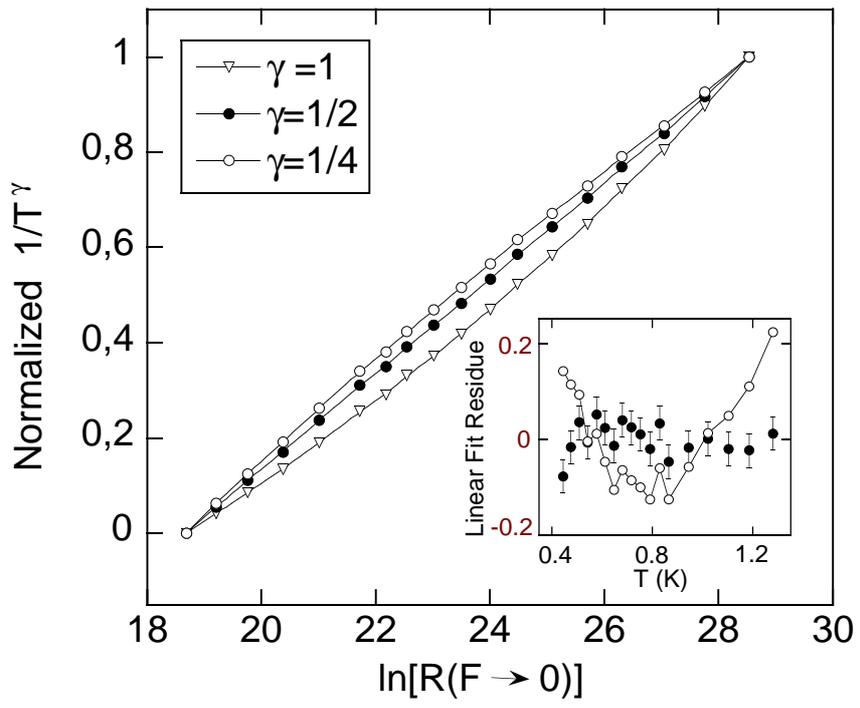
FIG. 4. Normalized
value of $T^{-\gamma}$ for 3 values of $\gamma$ 
as a function of the logarithm of the resistance at zero electric field. 
The normalized $T^{-\gamma}$ is
$(T^{-\gamma}-T^{-\gamma}_{Min})/(T^{-\gamma}_{Max}-T^{-\gamma}_{Min})$
where $T_{Min}$ and $T_{Max}$ are the two extreme
 values of the temperature $T$ selected for the plot.
Inset : residue (i.e. measured minus fitted value) of the linear fit 
of $\hbox{ln}[R(F \to 0)]$ as a function of $T^{-\gamma}$, for $\gamma=1/4$ 
(open circles) and $\gamma=1/4$ (closed circles).

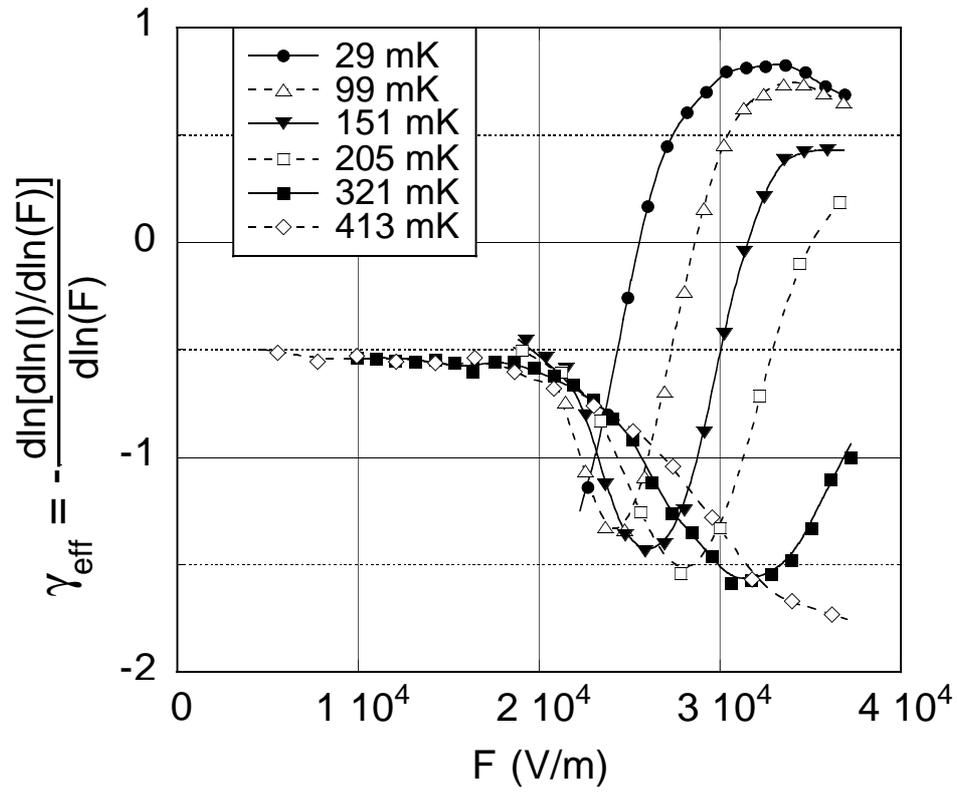
FIG. 5. Effective exponent of the current vs electric field law, as a function of the
electric field, for 6 temperatures. 
   
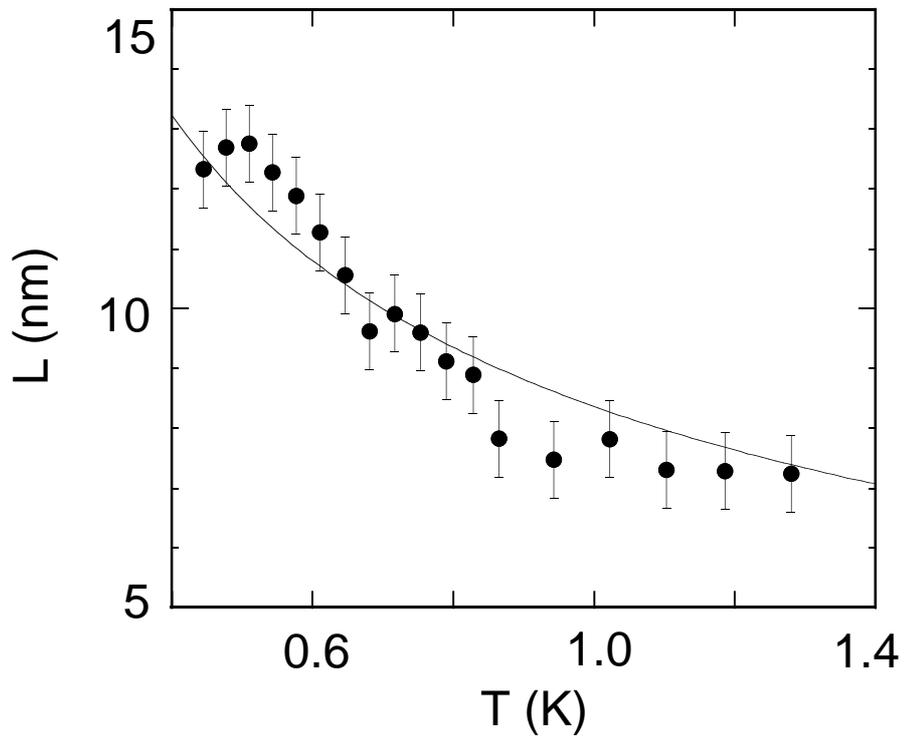
FIG. 6. The length $L$ extracted from the slopes
of the ln$(V/I)$ vs $F/T$ lines at $F \to 0$ using Eq. (4) (dots), and the fitted $L=L_0 \ (T_0/T)^{1/2}$ law (continuous line).

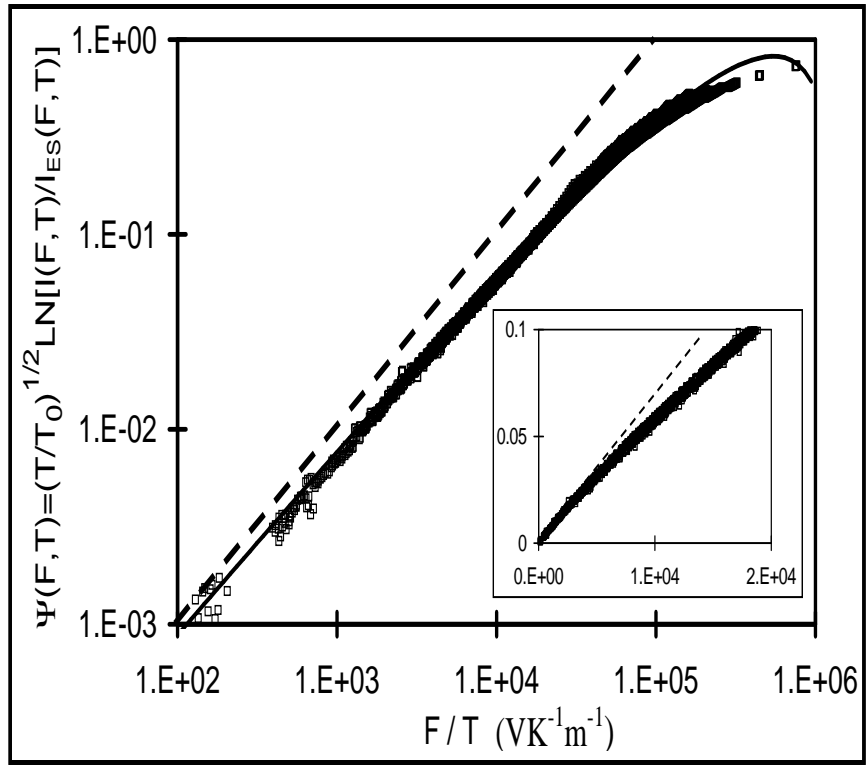
FIG. 7. The $\Psi(F,T)$ function (given by $(T/{T_{0}})^{1/2} \ln(I/I_{E.S.})$, with $I_{E.S.} = (V/R_0) \exp[-(T_0/T)^{1/2}]$)
 as a function of F/T for all the data except the 
very high field points. $\Psi$ increases more slowly than $F/T$ (the dashed line corresponds to 
$ \Psi \propto F/T $)
 and its evolution can be fitted (solid line) using the B\"ottger {\it et al.} prediction $\Psi = 
\Gamma (F/T) 
- \Delta (F/T)^{\beta}$ with 
 $\beta = 1.15 \pm 0.1$,
$\Gamma = (1.25 \pm 0.5) \times 10^{-5}\ \hbox{K} \, \hbox{m} \, \hbox{V}^{-1}$ and ${\Delta^{{1}/{\beta}}}/{\Gamma} = 0.70 \pm 0.07$. Inset : same $\Psi$ vs. $F/T$ dependence with linear scales showing that the $\Delta ({F}/{T})^{\beta}$ term plays a role at low ${F}/{T}$ values; the dashed line corresponding to $\Psi \propto F/T$. 

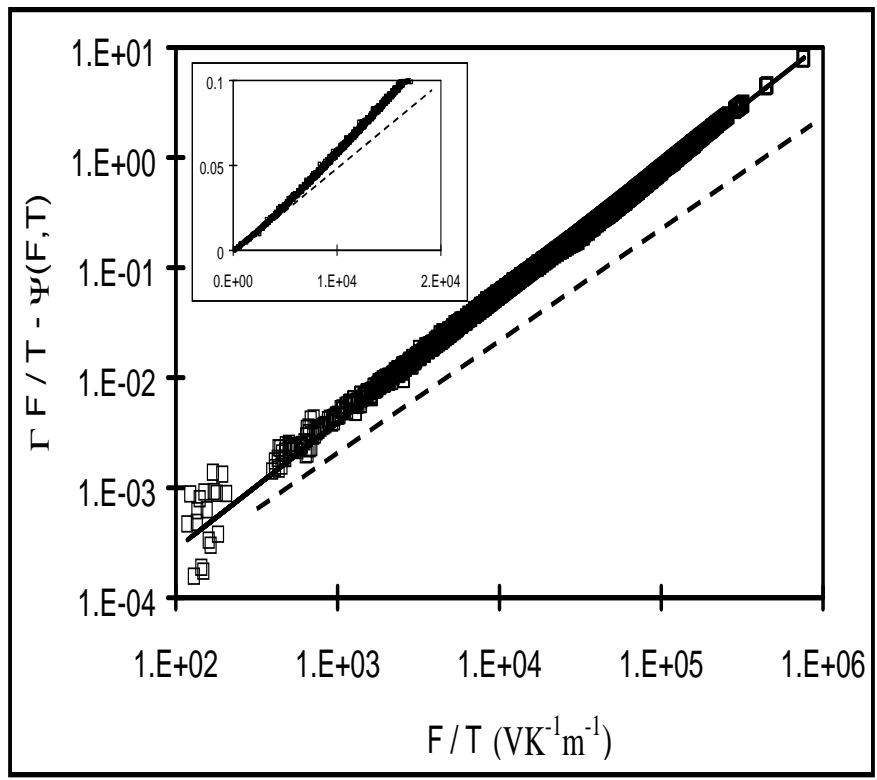
FIG. 8. $F/T$ dependence of  $\Gamma ({F}/{T}) - \Psi(F,T)$ , with $\Gamma = 1.25 \times 
10^{-5} \hbox{K} \, \hbox{m} \, \hbox{V}^{-1}$, for all the data points except the very high field ones.
The dashed line corresponds to a $\Delta (F/T)^{\beta}$ dependence with $\beta=1$, and shows that $\beta >1$. Solid line : fit with $\beta=1.15$ and $\Delta = 1.4 \times 10^{-6} (\hbox{Km/V})^{\beta}$. Inset : same data with linear scales, the dashed line corresponding to a $\sim F/T$ dependence.

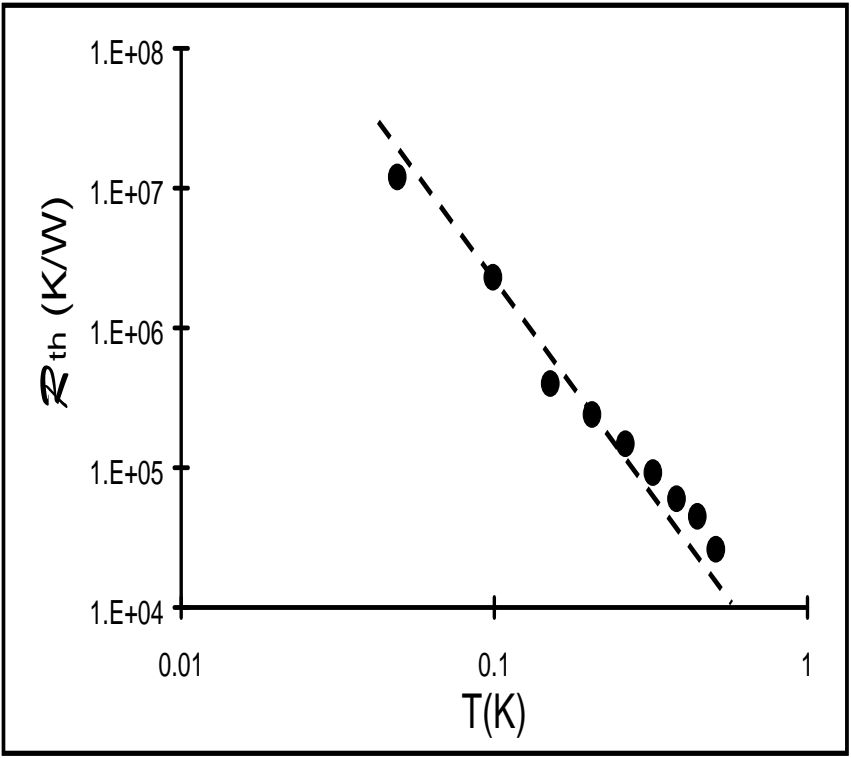
FIG. 9. Thermal resistance ${\cal R}_{th}(T)$ measured
by dissipating a controlled power in the small $1 \, \hbox{M}\Omega$ chip (see Fig. 1).
 Dashed line : fit of the data with 
a $T^{-3}$ law. 

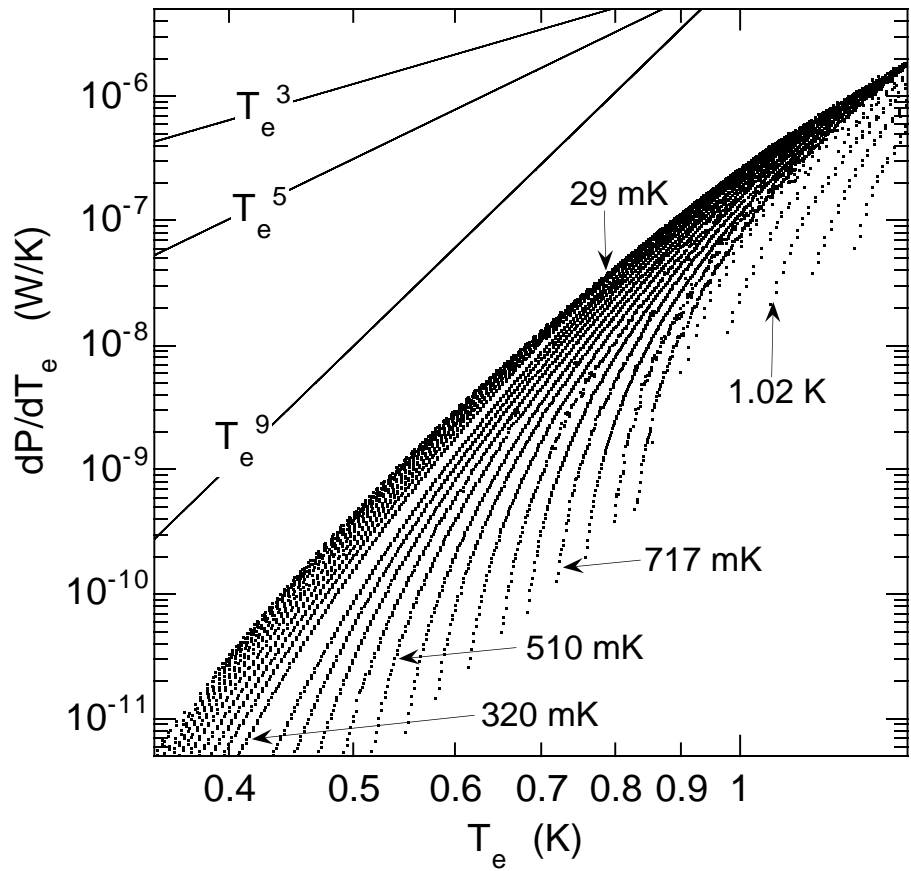
FIG. 10. Derivative of the power dissipated in the YSi sample
with respect to the effective electron temperature $T_e$, as 
a function of $T_e$. For each ($I,V$) point, $T_e$ is calculated
using Eq. (A1). The points line up in curves corresponding to different
cryostat temperatures among which 5 are indicated in the figure.
For comparison, the three straight lines correspond to 3 power laws
$dP/dT_e \propto T_e^{(\gamma - 1)}$ with $\gamma -1$ = 3, 5 and 9.


\begin{thebibliography}{99}


\bibitem{Anderson58} P.W. Anderson, Phys. Rev. {\bf 109}, 1492 (1958). 

\bibitem{Mott69} N.F. Mott, Phil.Mag. {\bf 19}, 835 (1969).

\bibitem{Ambegaokar71} V. Ambegaokar, B.I. Halperin, J.S. Langer, Phys. Rev. B {\bf 4}, 2612 (1971).

\bibitem{Pollak72} M. Pollak, J. Non Cryst. Sol. {\bf 11}, 1 (1972).

\bibitem{Shklovskii74} B.I. Shklovskii and A.L. Efros, {\it Electronic properties of 
doped Semiconductors}, Springer series in Solid-State Science Vol. 45 
(Springer, New-York, 1984)

\bibitem{Aharony92} A. Aharony, Y. Zhang, P. Sarachik, Phys. Rev. Lett. {\bf 68}, 3900 (1992).

\bibitem{Shklovskii72} B.I. Shklovskii, Sov. Phys. Semicond. {\bf 6}, 1964 (1973).

\bibitem{Apsley75} N. Apsley and H.P. Hughes, Phil. Mag. {\bf 30}, 963 (1974); 
 N. Apsley and H.P. Hughes, Phil. Mag. {\bf 31}, 1327 (1975).

\bibitem{Pollak76} M. Pollak and I. Riess, J. Phys. C {\bf 9}, 2339 (1976).

\bibitem{Shklovskii76} B. I. Shklovskii, Sov. Phys. Semicond. {\bf 10}, 855 (1976).

\bibitem{Rentzsch79} R. Rentzsch, I. S. Shlimak and H. Berger, Phys. Stat. Sol. (a) {\bf 54}, 487 (1979).

\bibitem{Vandermeer82} M. Van Der Meer, R. Schuchardt, R. Keiper, Phys. Stat. Sol. B
 {\bf 110}, 571 (1982).

\bibitem{Hill71} R.M. Hill, Phil. Mag. {\bf 24}, 1307 (1971).

\bibitem{Shklovskii75} B.I. Shklovskii, A.L. Efros, Usp. Fiz. Nauk. {\bf 117}, 401 (1975). 

\bibitem{Feng91} S. Feng, J.L. Pichard, Phys. Rev. Lett. {\bf 67}, 753 (1991)
 and references therein. 

\bibitem{Bottger80} H. B\"ottger and V.V. Bryksin, Phil. Mag. B {\bf 42}, No 2, 
297 (1980); H. B\"ottger and V.V. Bryksin, Phys. Stat. Sol. B 
{\bf 96}, 219 (1979).

\bibitem{Bottger85} H. B\"ottger, P. Szyler, D. Wegener, Phys. Stat. Sol. B
 {\bf 128}, K179 (1985).

\bibitem{Bottger86} H. B\"ottger, P. Szyler, D. Wegener, Phys. Stat. Sol. B
{\bf 133}, K143 (1986).

\bibitem{Talamantes87} J. Talamantes, M. Pollak, R. Baron, J. Non Cryst. Sol.
{\bf 97-98}, 555 (1987).

\bibitem{Levin84} E. I. Levin and B. I. Shklovskii, Sov. Phys. Semicond. {\bf 18}, 534 (1984).

\bibitem{Bottger82} H. B\"ottger and V.V. Bryksin, Phys. Stat. Sol. B {\bf 113},  
9 (1982).

\bibitem{Bottger84} H. B\"ottger and D. Wegener, Phys. Stat. Sol. B {\bf 121},  413 (1984); H. B\"ottger and 
V.V. Bryksin in {\it Hopping Conduction in Solids}, pp. 236-259,  
VCH Publishers (USA), ISBN 0-89573-481-8; H. B\"ottger and D. Wegener, in 
{\it Hopping and Related Phenomena}, 
p. 317, H. Fritzsche, M. Pollak eds., World Scientific Publishing Company, (1990).

\bibitem{Talamantes92} J. Talamantes and J. Floratos, Phil. Mag. B 
{\bf 65}, No 4, 627 (1992).

\bibitem{Kirkpatrick86} T.R. Kirkpatrick, Phys. Rev. B {\bf 33}, 780, (1986).

\bibitem{note1}  We have $E_{F} = E_{Mob}-k_{B}T_{1}$ (see Sec. III) with the mobility edge $E_{Mob}$ 
given by the Ioffe-Regel criterion $(2mE_{Mob})^{{1}/{2}}\lambda_{el} = \hbar$ where  $\lambda_{el}$
 is the elastic mean free path. Since our sample is amorphous, $\lambda_{el} \simeq 0.3$ nm) yielding 
$E_{Mob} \simeq 0.35 \, \hbox{eV} \simeq E_{F}$. With $\xi = 2-6 \, \hbox{nm}$ (see Sec. V) it is found 
that 
$e F \xi \le 10^{-4} \times  E_{F}$ even for the maximum field used $4\times 10^{4} \, \hbox{V/m}$. 

\bibitem{Morgan71} M. Morgan and P. A. Walley, Phil. Mag. {\bf 27}, 1151 (1971).

\bibitem{Telnic73} M. Telnic, L. Vescan, N. Croitoru and C. Popescu, Phys. Stat. Sol. (b)
{\bf 59}, 699 (1973).

\bibitem{Marshall73} J.M. Marshall and G.R. Miller, Phil. Mag. {\bf 27}, 1151 (1972).

\bibitem{Elliott74} P.J. Elliott, A.D. Yoffe and E.A. Davis, American Institute of Physics Conf.
Proc. {\bf 20}, 311 (1974).

\bibitem{Nair77} K. Nair and S.S. Mitra, J. Non-Crystalline Solids {\bf 24}, 1 (1977).

\bibitem{Aleshin87} A. N. Aleshin and I. S. Shlimak, Sov. Phys. Semicond. {\bf 21}, 289 (1987).

\bibitem{Dvurechenskii88} A. V. Dvurechenskii, V. A. Dravin and A. I. Yakimov, JETP Lett. {\bf 48}, 156 (1988).

\bibitem{Nebel92} C. E. Nebel, R. A. Street, N. M. Johnson and C. C. Tsai, Phys. Rev. B {\bf 46}, 6803 
(1992).

\bibitem{Popescu98} B. Popescu, T. Wright, C. J. Adkins and S. Iovan, Phys. Stat. Sol. B {\bf 205}, 77 
(1998).

\bibitem{Redfield75} D. Redfield, Adv. Phys. {\bf 24}, 463 (1975).

\bibitem{Zabrodskii80} A.G. Zabrodskii and I.S. Shlimak, Sov. Phys. Semicond. {\bf 11}, 430 (1980).

\bibitem{Rosenbaum80} T. F. Rosenbaum, K. Andres and G. A. Thomas, Solid State Comm. {\bf 35}, 663 
(1980).

\bibitem{Ionov87} A. N. Ionov, M. N. Matveev, I. S. Shlimak and R. Rentch, JETP Lett. {\bf 45}, 311 (1987).

\bibitem{Aladashvili90} D. I. Aladashvili, Z. A. Adamiya, K. G. Lavdovskii, E. I. Levin and B. I. Shklovskii, in 
{\it Hopping and Related Phenomena}, edited by H. Fritzsche and M. Pollak (1990) pp. 283-297,  World 
Scientific Publishing Company, (1990). 

\bibitem{Kenny89} T. W. Kenny, P. L. Richards, I. S. Park, E. E. Haller and J. W. Beeman, Phys. Rev. {\bf 
39}, 8476 (1989).

\bibitem{ChenGang89} Chen Gang, H. D. Koppen, R. W. van der Heijden, A. T. A. M. de Waele and H. M. 
Gijsman,  Solid State Comm. {\bf 72}, 173 (1989).

\bibitem{Timchenko89} I. N. Timchenko, V. A. Kasiyan, D. D. Nedeoglo and A. V. Simashkevich,  Sov. 
Phys. Semicond. {\bf 23}, 148 (1989).

\bibitem{Tremblay89} F. Tremblay, M. Pepper, R. Newbury, D. Ritchie, D. C. Peacock, J. E. F. Frost and 
G. A. C. Jones, Phys. Rev. B {\bf 40}, 3387 (1989).

\bibitem{Wang90} N. Wang, F.C. Wellstood, B. Sadoulet, E. E. Haller and J. Beeman, Phys. Rev. B {\bf 
41}, 3761 (1990).

\bibitem{Grannan92} S. M. Grannan, A. E. Lange, E. E. Haller, J. W. Beeman, Phys. Rev. B {\bf 45}, 4516 
(1992).

\bibitem{VanderHeijden92} R. W. Van der Heijden, G. Chen, A. T. A. M. de Waele, H. M. Gijsman
and F. P. B. Tielen, Phil. Mag. B  {\bf 65}, 849 (1992).

\bibitem{Stephanyi97} P. Stephanyi, C. C. Zammit, P. Fozooni, M. J. Lea and G. Ensell, J. Phys.: 
Condens. Matter {\bf 9}, 881 (1997).

\bibitem{LTD5} A. Alessandrello {\it et al.}, in {\it Proceedings of the 5th Int. Workshop on Low Temperature
Detectors}, edited by S. E. Labov and B. A. Young (Berkeley, July 29 - August 3 1993), J. Low 
Temperature
Physics {\bf 93} (3/4) (1993) pp. 207-212 and pp. 337-342; M. Frank {\it et al.} {\it ibid.}, pp. 213-218; E. 
Aubourg {\it et al.} {\it ibid.}, pp. 289-294; L. Dumoulin {\it et al.} {\it ibid.}, pp. 301-306; N. Perrin {\it et al.} 
{\it ibid.}, pp. 313-317; J. Soudee {\it et al.} {\it ibid.}, pp. 319-324; K. Djotni {\it et al.} {\it ibid.}, pp. 325-
329; X. X. Wang {\it et al.} {\it ibid.}, pp. 349-354.     
    
\bibitem{Alessandrello95}  A. Alessandrello {\it et al.}, in {\it Proceedings of the 6th Int. Workshop
 on Low Temperature Detectors}, edited by H.R. Ott and A. Zehnder (Beatenberg, Switzerland 28 August - 1 
September 1995), Nucl. Inst. and Meth. {\bf A} 370 (1995), pp. 244 - 246.  

\bibitem{LTD7} A. Alessandrello {\it et al.}, in {\it Proceedings of the 7th Int. Workshop on Low Temperature
Detectors}, edited by S. Cooper (Munich, 27 July - 2 August 1997) pp. 138-141;
 S. M. Grannan and P. L. Richards {\it ibid.}, pp. 160-161.  

\bibitem{Yvon96} D. Yvon, L. Berg\'e, L. Dumoulin, P. De Marcillac, S. Marnieros, P. Pari, G. Chardin, 
Nucl. Inst. and Meth. A {\bf 370}, 201 (1996).

\bibitem{Dumoulin96} L. Dumoulin, L. Berg\'e, S. Marnieros, J. Lesueur, Nucl. Inst. and Meth. A {\bf 370}, 
211 (1996).

\bibitem{Perrin97} N. Perrin, J. Appl. Phys. {\bf 82}, 3341 (1997).


\bibitem{Boucher88} M. Sanquer, R. Tourbot, B. Boucher, Europhys. Lett. {\bf 7}, 
635 (1988).

\bibitem{Pichard90} J.L. Pichard, M. Sanquer, K. Slevin, P. Debray, Phys. Rev. Lett.
 {\bf 65}, 1812 (1990).

\bibitem{Hernandez92} P. Hernandez and M. Sanquer, Phys. Rev. Lett. {\bf 68}, 
1402 (1992).

\bibitem{Hernandez94} P. Hernandez, F. Ladieu, M. Sanquer, D. Mailly, Physica B 
{\bf 194-196}, 1141 (1994).

\bibitem{Specht95} M. Specht, L.P. Levy, F. Ladieu, M. Sanquer, Phys. Rev. Lett.
{\bf 75}, 3902 (1995).

\bibitem{Ladieu96} F. Ladieu, M. Sanquer, J.P. Bouchaud, Phys. Rev. B {\bf 53}, 
No 3, 973 (1996).

\bibitem{Sze81} S.M. Sze, {\it Physics of Semiconductor Devices},
 (Wiley, New York, 1981), p. 403.
 
\bibitem{note2} The choice of $\Psi$ was also motivated by the fact that we first 
 tried to analyze our $I(F,T)$ dependence by using $\Phi (F,T)  = \sqrt{(T/T_0)} \ln{[(I(F,T)-I_{E.S.})/I_1]} 
$, expanding  
on the intuition that the current $I(F,T)$ might be the
 {\it sum} of the linear current $I_{E.S.}(F,T)$ deduced from Eq. (1)
 and of a non linear current given by Eq. (3). This failed since $\Phi$
 appears to {\it be negative} at low  
 ${F}/{T}$, and particularly $\Phi (F\to 0, T) \ll 0$.
 Very importantly, when ${F}/{T} \ge 3 \times 10^{4} \, \hbox{V} \ \hbox{K}^{-1} \hbox{m}^{-1}$, $I_{1}$ can 
be chosen such that  $\Phi(F,T) \simeq \Psi(F,T)$. The values of $\alpha, \beta, A,B$ 
obtained from a fit of $\Psi$ are thus consistent with those from a fit of 
$\Phi({F}/{T})$ for $F/T \ge 3 \times 10^{4} \hbox{V} \ \hbox{K}^{-1} \hbox{m}^{-1}$.

\bibitem{Imry82} Y. Imry, Y. Gefen, D.J. Bergman, Phys. Rev. B {\bf 26}, 3436 (1982).

\bibitem{Schwartz89} E.T. Schwartz and R.O. Pohl, Rev. Mod. Phys. {\bf 61}, 605 (1989).

\bibitem{Cooper87} J. R. Cooper, W. P. Beyermann, S. W. Cheong and G. Gruner, Phys. Rev. B {\bf 36}, 
7748 (1987).

\bibitem{Abrahams86} E. Abrahams, P.A. Lee, Phys. Rev. B {\bf 33}, 683, (1986).

\bibitem{Nakhmedov87} E.P. Nakhmedov, V.N. Prigodin, Y.A. Firtov, Sov. JETP {\bf 65}, 1202, (1987); P. 
Sebbah, D. Sornette, C. Vanneste, Phys. Rev. B {\bf 48}, 12506 (1993).

\bibitem{Ortuno98} A. Prez-Garrido, M. Ortuno, E. Cuevas, J. Ruiz, M. Pollak, 
Phys. Rev. B {\bf 55}, R8630 (1997); J. Talamantes and A. Moebius, Phys. Stat. Sol. B {\bf 205}, 45 
(1998).

\end{thebibliography}
\end{document}